\documentclass[a4paper,10pOAt,twocolumn,amsmath,amssymb,superscriOAptaddress,floatfix,nofootOAinbib]{revtex4-1}
\usepackage{mathrsfs,bm}
\usepackage{longtable}
\usepackage{txfonts}
\usepackage{amssymb}
\usepackage{indentfirst}
\usepackage{graphicx,,booktabs}
\usepackage{multirow}
\usepackage{color}

\begin{document}

\title{Neutron structure function via a maximum entropy analysis}

\author{Chengdong Han}
\email{chdhan@impcas.ac.cn}
\affiliation{Institute of Modern Physics, Chinese Academy of Sciences, Lanzhou 730000, China}
\affiliation{State Key Laboratory of Heavy Ion Science and Technology, Institute of Modern Physics, Chinese Academy of Sciences, Lanzhou 730000, China}
\affiliation{School of Nuclear Science and Technology, University of Chinese Academy of Sciences, Beijing 100049, China}

\author{Rong Wang}
\email{rwang@impcas.ac.cn (Corresponding Author)}
\affiliation{Institute of Modern Physics, Chinese Academy of Sciences, Lanzhou 730000, China}
\affiliation{State Key Laboratory of Heavy Ion Science and Technology, Institute of Modern Physics, Chinese Academy of Sciences, Lanzhou 730000, China}
\affiliation{School of Nuclear Science and Technology, University of Chinese Academy of Sciences, Beijing 100049, China}

\author{Xurong Chen}
\email{xchen@impcas.ac.cn (Corresponding Author)}
\affiliation{Institute of Modern Physics, Chinese Academy of Sciences, Lanzhou 730000, China}
\affiliation{State Key Laboratory of Heavy Ion Science and Technology, Institute of Modern Physics, Chinese Academy of Sciences, Lanzhou 730000, China}
\affiliation{School of Nuclear Science and Technology, University of Chinese Academy of Sciences, Beijing 100049, China}

\begin{abstract}
  We employ the maximum entropy method to extract the valence quark distributions of the neutron at a low scale, \( Q_0^2 \).
  At this initial scale, the neutron is defined to contain only three valence quarks, with no contributions from sea quarks or gluons.
  The distributions of these initial valence quarks are constrained by principles from quark models, quark-hadron duality, and quark confinement.
  Employing the DGLAP equations supplemented by parton-parton recombination corrections, we derive the neutron structure function \( F_2^{n} \) at higher scales \( Q^2 \).
  The resulting ratio of the neutron to proton structure functions, $F_2^{n}$/$F_{2}^{p}$, aligns well
  with the world deep inelastic scattering data at Bjoken variable $x<0.7$, particularly when accounting for uncertainties from model-dependent corrections.
  Notably, this ratio is in agreement with the JLab MARATHON data after considering the quark-hadron duality assumption, especially in the region of $x \gtrsim 0.7$.
  Additionally, our findings for $F_2^{n}$/$F_{2}^{p}$ correspond well with the JLab BONuS experimental results after considering the impact of nucleon resonance contamination
  in the region $x \gtrsim 0.4, 0.5, 0.6$. We further compare our predictions for $F_2^{n}$/$F_{2}^{p}$ and the \( u/d \) ratios in the limit as \( x \rightarrow 1 \) with
  existing theoretical calculations. Finally, we observe a minor violation of isospin symmetry between the proton and neutron, evidenced by the differences in valence quark distributions
  and the first-order moments of these distributions.
  % keywords: Neutron; Structure function $F_2$; Quark model; Quark-hadron duality; Valence quark distributions; Maximum entropy method; DGLAP equations with nolinear corrections
\end{abstract}

\maketitle

\section{Introduction}
\label{introduction}
The determination of the neutron structure function \( F_2^{n} \) \cite{NewMuon:1991exl,Arrington:2008zh,Weinstein:2010rt,Arrington:2011qt, CLAS:2011qvj, Hen:2011ad, CLAS:2014jvt, Niculescu:2015wka, Accardi:2016qay, Szumila-Vance:2020zpt} is essential for advancing our understanding of the quark structure within the nucleon.
The relatively limited knowledge of neutron structure information, in contrast to that of the proton, is primarily due to the scarcity of neutron targets,
making the study of \( F_2^{n} \) a topic of significant interest.
Currently, most of the available neutron structure function information is extracted from a comprehensive review of global deep inelastic scattering (DIS) data
obtained from proton, deuteron, tritium, and \( ^3\text{He} \) targets throughout the full kinematic range. This extraction process also incorporates various nuclear correction models.

Recent advancements have emerged from the JLab MARATHON experiment \cite{JeffersonLabHallATritium:2021usd},
which utilized two high-resolution spectrometers for electron detection to measure the ratio of nucleon \( F_{2} \) structure functions,
denoted as \( F_{2}^{n}/F_{2}^{p} \), through deep inelastic electron scattering off \( ^3\text{H} \) and \( ^3\text{He} \) nuclei.
A novel analytical technique exploiting the mirror symmetry of \( ^3\text{H} \) and \( ^3\text{He} \) significantly mitigated theoretical uncertainties associated with this extraction.
The reported \( F_{2}^{n}/F_{2}^{p} \) ratio, spanning the Bjorken variable range \( 0.19 < x < 0.83 \), constitutes a notable improvement relative to prior measurements
conducted by SLAC and JLab. Additionally, the Barely Off-shell Nucleon Structure (BONuS) experiment at JLab \cite{Fenker:2008zz,CLAS:2011qvj,CLAS:2014jvt} facilitated
the measurement of quasi-free neutron structure function data across both nucleon-resonance and DIS regions by detecting low-momentum spectator protons at backward angles
during semi-inclusive scattering with deuterons. 

Over the past few decades, considerable interest has also been directed toward the ratio \( F_{2}^{n}/F_{2}^{p} \) in the large Bjorken variable \( x \) region.
The neutron structure function \( F_{2}^{n} \) \cite{Li:2023yda} and the corresponding \( F_{2}^{n}/F_{2}^{p} \) ratio are regarded as pivotal
for determining the \( d/u \) (or \( u/d \) for the neutron) ratio in the limit as \( x \) approaches 1.

It is important to note that the analysis results of neutron structure functions obtained thus far do not stem from an authentic free neutron target. The challenges posed by the short lifetime of the neutron and the low energy and intensity of existing neutron beams have rendered direct experimental investigations of its internal structure impractical. Consequently, information regarding neutron structure is typically inferred from deuterium or \( ^3\text{He} \) target data obtained via DIS measurements. The neutron structure function is subsequently deduced by applying corrections based on known proton structure functions, which necessitate model-dependent adjustments to account for nuclear binding, Fermi motion, EMC effects, final state interactions, and nucleon off-shell effects \cite{Whitlow:1991uw, Frankfurt:1988nt, Gomez:1993ri, Melnitchouk:1994rv, Melnitchouk:1995fc}. Despite these corrections, the extraction of the neutron structure function \( F_{2}^{n} \), especially in the large \( x \) domain (\( x > 0.8 \)), remains fraught with uncertainties. Varied nuclear correction models yield divergent theoretical predictions for the \( F_{2}^{n}/F_{2}^{p} \) and \( d/u \) (or \( u/d \) for the neutron) ratios in the limit as \( x \) approaches 1 \cite{Weinstein:2010rt, Hen:2011ad, Nakano:1991kh, Holt:2010vj}.

In this work, we seek to determine the initial valence quark distributions at a scale \( Q^{2}_{0} \) for the neutron using the Maximum Entropy Method (MEM). This approach leverages existing structure information and properties of the neutron within the frameworks of quark models and quantum chromodynamics (QCD). The MEM has been effectively applied to investigate parton distributions of protons \cite{Wang:2014lua, Han:2016bsw}, mesons \cite{Han:2018wsw, Han:2020vjp}, and exotic hadrons such as the \( Z_{c}(3900) \) \cite{Han:2024xjg}. By employing DGLAP equations with nonlinear corrections \cite{Chen:2013nga, Wang:2016mzo, Wang:2016sfq, Chen:2014nba} to evolve the nonperturbative neutron input obtained via MEM to higher scales, we will compare the resulting neutron structure function and \( F_{2}^{n}/F_{2}^{p} \) ratio with available experimental data. Furthermore, we will discuss our findings regarding the structure function ratio \( F_{2}^{n}/F_{2}^{p} \) and the \( u/d \) ratio for the neutron, contrasting them with predictions from various theoretical models in the limit as \( x \) approaches 1. Finally, we will compute the valence quark momentum distribution differences between protons and neutrons as functions of the Bjorken variable \( x \), along with the first-order moments differences as functions of \( Q^{2} \), to assess the extent to which protons and neutrons adhere to isospin symmetry.

\section{Determining the nonperturbative input by maximum entropy method}
\label{SecII:determine initial PDFs}
Under certain constraints, the MEM principle can determine a reasonable initial valence quark distributions inside neutron,
including constraints on the quark model, quark-hadron duality and quark confinement.

\subsection{Quark model constraints}
The definition for quark model is a classification scheme that expounds the quantum numbers of the baryons and mesons by assuming
that baryons are composed of three valence quarks and mesons are composed of a pair of quark and anti-quark.
The parton distribution functions (PDFs) of nucleon at high $Q^2$ obtained by performing DGLAP equations with parton-parton recombinations
depend on the initial parton distributions at low scale $Q^2_0$.
In this analysis, a naive nonperturbative input of the neutron only contains three valence quarks without other nonperturbative components (sea quarks and gluons),
which is the simplest initial parton distribution input \cite{Parisi:1976fz,Vainshtein:1976kd,Gluck:1977ah,Chen:2013nga}.
In the view of dynamical PDF model, the sea quarks and gluons are radiatively generated from three valence quarks of neutron at
high scale $Q^{2} > Q^{2}_{0}$.
We take the input scale $Q_0^2 = 0.067$ GeV$^2$ that is determined by the global QCD analysis of
experimental data of proton \cite{Wang:2016sfq}.
The running strong coupling and parton-parton correlation length $R$ which characterizes the strength
parton-parton recombination corrections are determined by a large number of DIS experimental data
at high $Q^{2}$ \cite{Gluck:1998xa,Wang:2016sfq}, which are the same as in the paper of valence quark distributions
of the proton from MEM \cite{Wang:2014lua}.

According to the framework of quark model, we have the valence sum rules and the momentum sum rule as constraints:
\begin{equation}
\int_0^1 u_v(x,Q_0^2)dx = 1,
\int_0^1 d_v(x,Q_0^2)dx = 2.
\label{ValenceSum}
\end{equation}
\begin{equation}
\int_0^1 x[u_v(x,Q_0^2)+d_v(x,Q_0^2)]dx = 1.
\label{MomentumSum}
\end{equation}

\subsection{Quark-hadron duality}
The most general function form to approximate valence quark distribution of nucleon is the time-honored canonical parametrization
$f(x)=Ax^B(1-x)^C$ \cite{Pumplin:2002vw}.
%The behavior of elastic scattering is closely related to that of deep-inelastic electron-nucleon scattering, particularly about duality.
Duality provides a unifying framework that links the behavior of elastic scattering and the electroproduction of nucleon resonances with that of deep-inelastic electron–nucleon scattering.
Within this context, ideas originating from strong-interaction processes suggest that a significant part of the observed features of inelastic electron–nucleon scattering
can be attributed to the nondiffractive component of virtual photon–nucleon scattering \cite{Bloom:1971ye}.
The correlation between resonances at low energies and non-Pomeranchukon exchanges,
characterized by falling total cross sections at high energies, forms part of the general concept of duality and acquires a quantitative formulation
through finite-energy sum rules \cite{Bloom:1970xb}.
Through finite-energy sum rules, Bloom and Gilman \cite{Bloom:1971ye} derived the quantitative relations between the elastic scattering form factors
and the deep-inelastic scattering and gave the ratio of the neutron structure function to the proton structure function
under the assumption the ``scaling" of the elastic form factors
as $x \rightarrow 1$ and $q^{2} \rightarrow \infty$, as
\begin{equation}
F_{2}^{n}/F_{2}^{p} \sim (\mu_{n}/\mu_{p})^{2}=0.47=K,
\label{ratio_K}
\end{equation}
where the $\mu_{n}$ and $\mu_{p}$ up are the magnetic moments of the neutron and proton, respectively.
After considering the quark-hadron duality mentioned above, the parametrization of nonperturbative input for the neutron in this analysis is as follows,
\begin{equation}
\begin{aligned}
  u_v(x,Q_0^2) = &\; A_u x^{B_u}(1-x)^{C_u} 
  \big[1 + D_u(1+x)^{C_u} \\
  &\; + E_{u}(1-x)^{2C_u} \big], \\
  d_v(x,Q_0^2) = &\; (4K)A_d x^{B_d}(1-x)^{C_d} 
  \big[1 + D_d(1+x)^{C_d} \\
  &\; + E_{d}(1-x)^{2C_d} \big].
\end{aligned}
\label{Parametrization}
\end{equation}
From this parameterization of the neutron valence quark, it can be seen that quark-hadron duality is automatically satisfied when $x$=1.
Furthermore, considering the isospin symmetry of proton and neutron, the $C_{u}=C_{d}^{p}$=2.456, $C_{d}=C_{u}^{p}$=1.000,
$A_d=A_{u}^{p}$=4.589 here \cite{Wang:2014lua}.

\subsection{Quark confinement}
The quark confinement is a typical feature of strong interaction in Non-Abelian gauge field theory \cite{Wilson:1974sk},
which means all quarks are confined in a small region of hadron size.
In this work, we use the Heisenberg uncertainty principle Eq.~\ref{Uncertainty} as the constraint \cite{Wang:2014lua},
\begin{equation}
\sigma_X\sigma_P \ge \frac{\hbar}{2},
\label{Uncertainty}
\end{equation}
where the $\sigma_{X}$ is the standard deviation of the spacial distribution of one valence quark in the neutron,
and the $\sigma_{P}$ is the standard deviation of the valence quark momentum distribution in the neutron.
Here, $\sigma_X$ is directly related to the magnetic radius ($R_{m}$) of the neutron, which is 0.864 fm \cite{zyla202079}.
A simple estimation of $\sigma_X$ is $\sigma_X = (2\pi R_{m}^3/3)/(\pi R_{m}^2) = 2R_{m}/3$,
for the neutron that is approximately spherical in shape.
In addition, $\sigma_X$ of each $d$ valence quark is divided by $2^{1/3}$ as there are two $d$ valence quarks in the neutron.
Then we get $\sigma_{X_u}=2R_{m}/3$ and $\sigma_{X_d} = 2R_{m}/(3\times 2^{1/3})$
for $u$ and $d$ valence quarks inside neutron respectively \cite{Wang:2014lua,Han:2016bsw}.
Furthermore, the standard deviation of momentum fraction $x$ at initial scale $Q_0^2$ is defined as follows,
\begin{equation}
\sigma_x = \frac{\sigma_P}{M},
\label{xDeviation}
\end{equation}
where $M$ is the rest mass of neutron \cite{zyla202079}.
Finally, the constraints for valence quark distributions from QCD color confinement
and Heisenberg uncertainty principle \cite{Wang:2014lua} are written as,
\begin{equation}
\begin{aligned}
&\sqrt{\left<x_u^2\right>-\left<x_u\right>^2} = \sigma_{x_u}, &\sqrt{\left<x_d^2\right>-\left<x_d\right>^2} = \sigma_{x_d}\\
&\left<x_u\right> = \int_0^1 xu_v(x,Q_0^2)dx, &\left<x_d\right> = \int_0^1 x\frac{d_v(x,Q_0^2)}{2}dx\\
&\left<x_u^2\right> = \int_0^1 x^2u_v(x,Q_0^2)dx, &\left<x_d^2\right> = \int_0^1 x^2\frac{d_v(x,Q_0^2)}{2}dx.\\
\end{aligned}
\label{xDeviation}
\end{equation}

\subsection{Maximum entropy method}
\begin{figure}[htp]
  \begin{center}
    \includegraphics[width=0.45\textwidth]{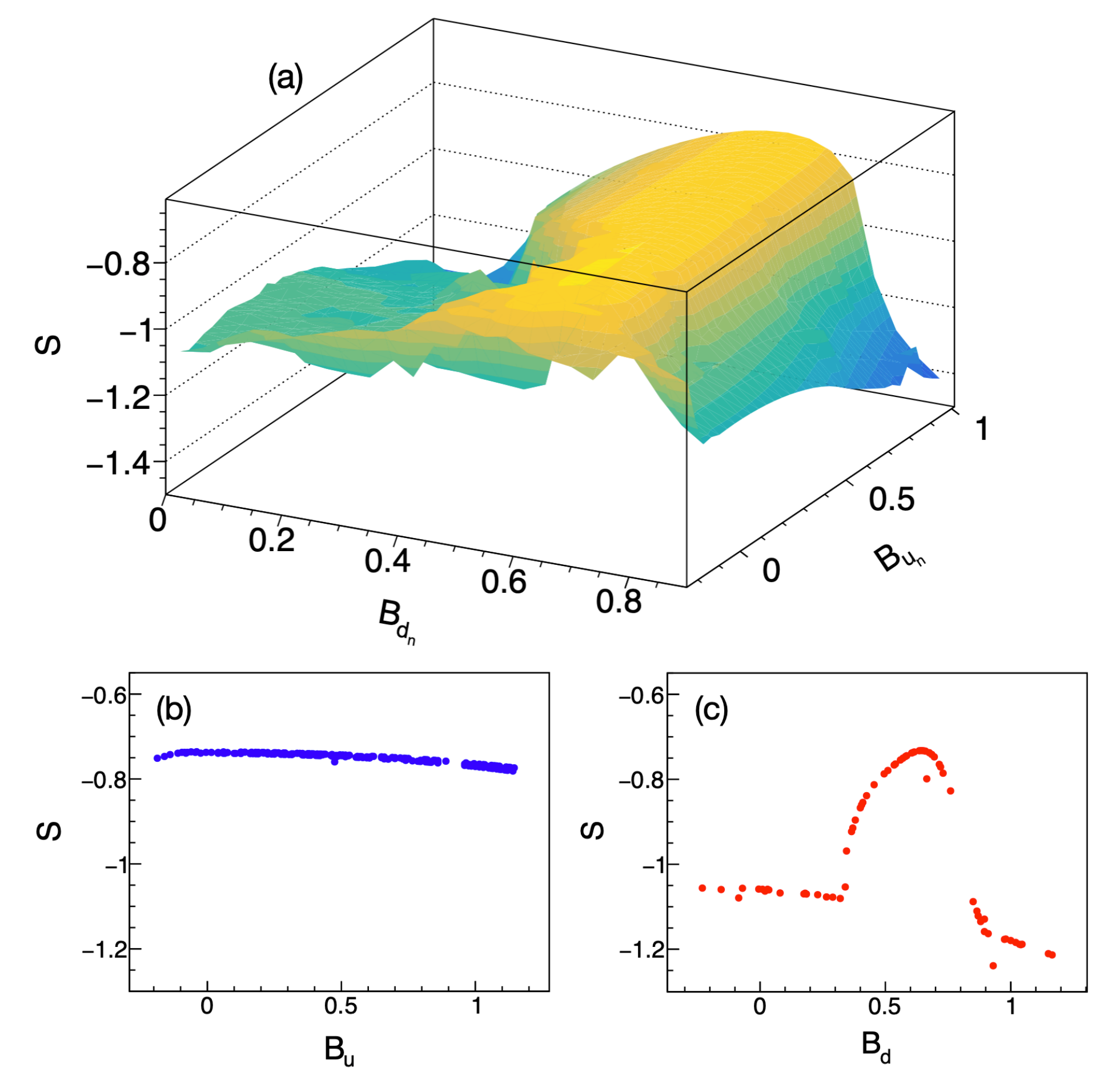}
    \caption{
      Entropy $S$ of valence quark distributions of neutron at $Q^{2}_{0}$ is plotted as a function of the free parameter $B_{u}$ and $B_{d}$,
      where subfigures (b) and (c) are projections of subfigure (a).
    }
    \label{fig:entropy}
  \end{center}
\end{figure}

According to the constraints (Eqs.~(1), (2), (3), (5), (6) and (7)) already introduced above,
there is only two unknown parameters $B_{u}$ and $B_{d}$ for Eq.~\ref{Parametrization}. 
In this work, these two unknown parameters, $B_u$ and $B_d$, are determined numerically.
By applying MEM, one can determine the reasonable valence quark distributions
under these constraints.
The generalized information entropy of valence quark distributions for neutron is given by,
\begin{equation}
\begin{aligned}
S=&-\int_0^1 \left[ u_v(x,Q_0^2){Ln}(u_v(x,Q_0^2)) \right. \\
  &+ \left.2 \frac{d_v(x,Q_0^2)}{2}{Ln}\left(\frac{d_v(x,Q_0^2)}{2}\right)\right] dx.
\end{aligned}
\label{EntropyDefinition}
\end{equation}
When the entropy $S$ of the entire neutron system reaches its maximum value, the values of the two unknown parameters under the given constraints can be determined,
and the optimal parameterized initial valence quark distributions can be further obtained.
Figure~\ref{fig:entropy} shows the information entropy $S$ of the valence quarks at the input scale $Q^{2}_{0}$
as a function of the free parameter $B_u$ and $B_{d}$.
Entropy $S$ value peaks at $B_u$=-0.096, $B_{d}$=0.645.
Hence the corresponding valence quark distributions from MEM are given by,

\begin{equation}
  \begin{aligned}
    u_v(x,Q_0^2) = &\; 5.673x^{-0.096}(1-x)^{2.456} \big[1-0.631(1-x)^{2.456} \\
    &\; - 0.251(1-x)^{4.912}], \\
    d_v(x,Q_0^2) = &\; 4.589x^{0.645}(1-x)^{1.000} \big[1-2.774(1-x)^{1.000} \\
    &\; + 4.320(1-x)^{2.000}].
  \end{aligned}
  \label{InitialValence}
\end{equation}

\section{Results and discussions}
\label{Results and discussions}

\subsection{PDFs and structure functions of neutron}
\begin{figure}[htp]
  \begin{center}
    \includegraphics[width=0.45\textwidth]{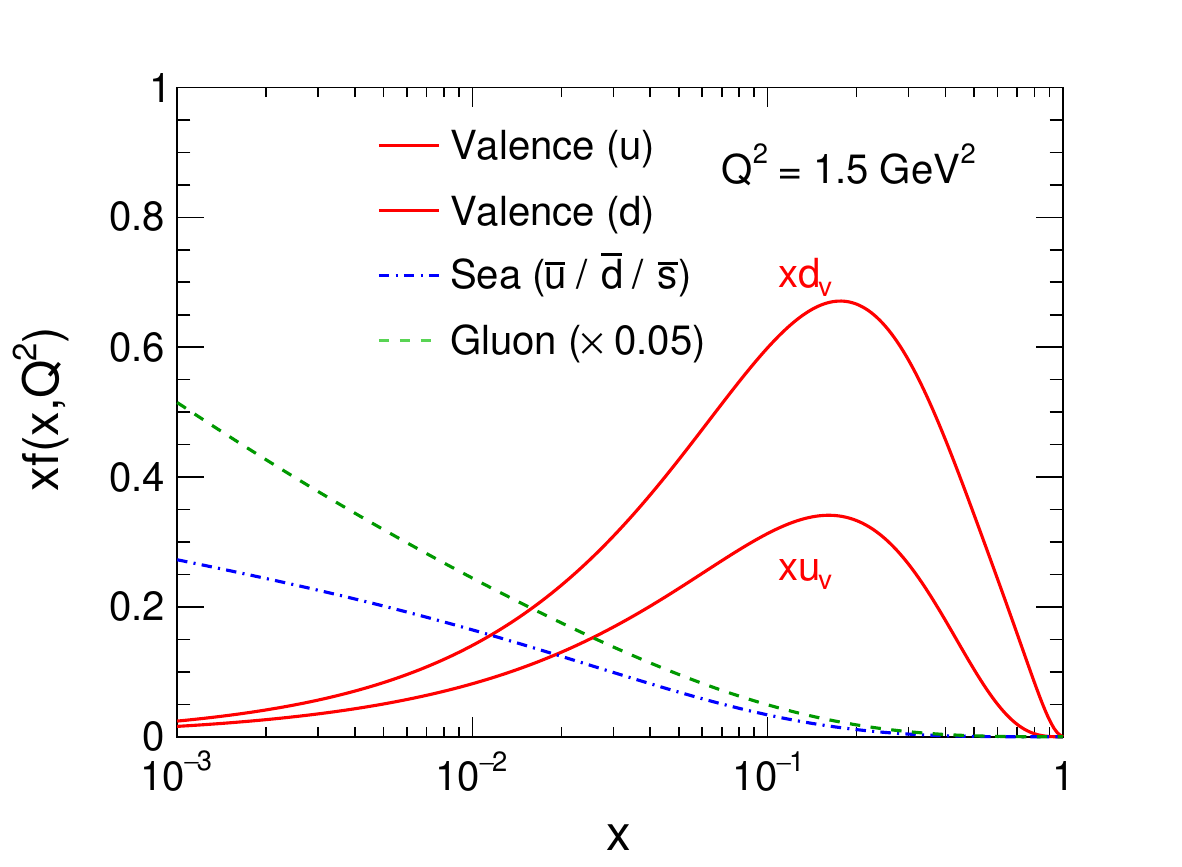}
    \caption{
      The predicted valence quark, sea quark and gluon distributions of neutron at $Q^{2} = 1.5$ GeV$^{2}$
      by performing DGLAP evolution equations with the parton-parton recombination corrections
      to the pure valence quark nonperturbative input from MEM.
    }
    \label{fig:neutron-PDFs}
  \end{center}
\end{figure}

Parton distribution functions of neutron are evaluated dynamically starting from
the obtained three valence quark input at the low scale $Q^{2}_{0}$.
By performing DGLAP equations \cite{Dokshitzer:1977sg, Gribov:1972ri, Altarelli:1977zs} with parton-parton recombination
corrections \cite{Chen:2013nga,Wang:2016mzo,Wang:2016sfq,Chen:2014nba},
the parton distributions at an arbitrarily high scale $Q^{2}$ can be determined with the nonperturbative input (Eq.~\ref{InitialValence}).
Figure~\ref{fig:neutron-PDFs} shows the predicted momentum distributions $xf_{i}(x,Q^{2})$ of up and down valence quarks, sea quarks and gluon of neutron,
  at $Q^2 = 1.5$ GeV$^2$. Here, $f_{i}(x,Q^{2})$ denotes the parton distribution function of quark or gluon flavor $i$, where $x$ is the Bjorken variable and $Q^{2}$
  is the probing scale.

The structure functions of nucleon reflect the characteristics of QCD defined by the asymptotic freedom of short
distances and the confinement of quarks on the long distance scale.
The unpolarized structure function $F_2(x)$ is directly related to the quark distribution functions.
According to the Quark-Parton model, the structure function \cite{Callan:1969uq} is written as,
\begin{equation}
2xF_{1}(x) = F_{2}(x) = \sum_{i} e_{i}^{2}xf_{i}(x),
\label{F2}
\end{equation}
where the subscript $i$ is flavor index, $e_{i}$ is the electrical
charge of the quark flavor $i$, and $xf_{i}(x)$ is the momentum fraction distribution of the quark of flavor $i$.
Since valence quarks dominate in the large $x$ region ($x > 0.1$), the $F_2(x)$ at large $x$ mainly
comes from the contribution of valence quarks.
In the small $x$ region ($x < 0.1$), the sea quarks begin to make an important contribution
to the structure function $F_2(x)$.

\begin{figure}[htp]
  \begin{center}
    \includegraphics[width=0.45\textwidth]{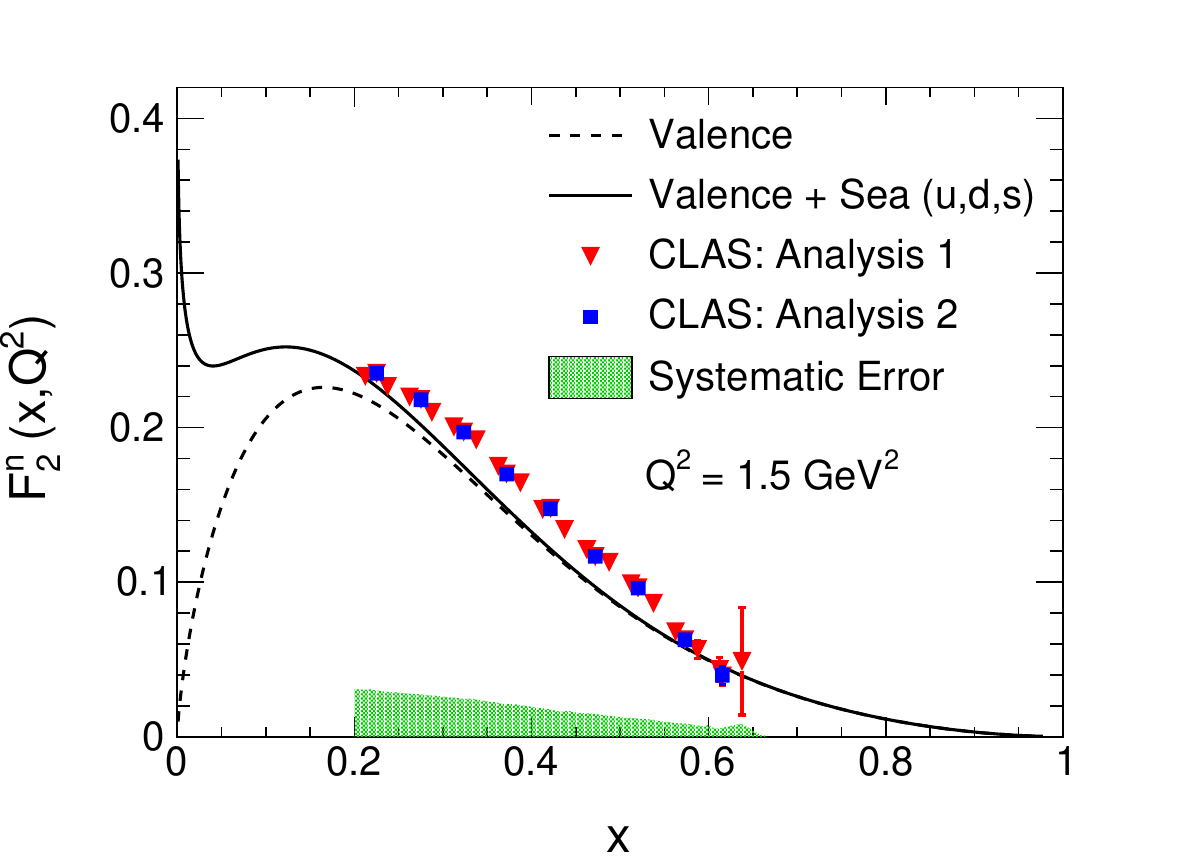}
    \caption{
      The predicted structure function of neutron $F_{2}^{n}$ as a function of Bjorken variable $x$
      compared with the BONuS experiment \cite{CLAS:2011qvj,CLAS:2014jvt}, where Analysis 1 (triangles) of BONuS experiment performs the Monte Carlo method
      and the Analysis 2 (squares) performs the ratio method.
      The solid curve includes the contribution of the sea quarks and the valence quarks,
      while the dashed curve includes only the valence quark contribution.
      The systematic uncertainties of the Monte Carlo method in $F_2^{n}$ extraction are shown as the green shaded band.
    }
    \label{F2n&BONuS}
  \end{center}
\end{figure}

The BONuS experiment at JLab \cite{CLAS:2011qvj,CLAS:2014jvt}
is via tagging very low momentum spectaor protons to measure the $F_{2}^{n}$ of the nearly free neutron
from the semi-inclusive scattering of electron off the deuterium.
This experimental method minimizes the off-shell effect significantly and reduces the nuclear binding uncertainties
by picking the spectator protons of momentum below 100 MeV/c and backward angles greater than $100$ degrees.
These selections ensure that the scattering occurs on the nearly free neutrons.
The current $F^{n}_{2}$ data collection cover the nucleon-resonance and deep-inelastic regions with a
wide range of Bjorken variable $x$ under $0.65 < Q^{2} < 4.52$ GeV$^{2}$.
Figure~\ref{F2n&BONuS} presents our determined $F_2^{n}$ as a function of $x$ compared with BONuS measurements,
where the dashed curve only includes the contribution of valence quarks and the solid curve includes the sum of valence quarks
and sea quarks.
It's worth emphasizing that at the input scale, the sea quark distributions are all zero. In the dynamical parton approach,
  all sea quarks at small $x$ are generated by the dynamic QCD evolution processes. That is to say, parton radiation is the dynamical origin of sea quarks inside the neutron.
The BONuS experiment provides two analyses: Analysis 1 based on the Monte Carlo method, and Analysis 2 based on the ratio method.
By comparisons, our MEM predictions are consistent with the BONuS data within the systematic uncertainties.
\begin{figure}[htp]
  \begin{center}
    \includegraphics[width=0.45\textwidth]{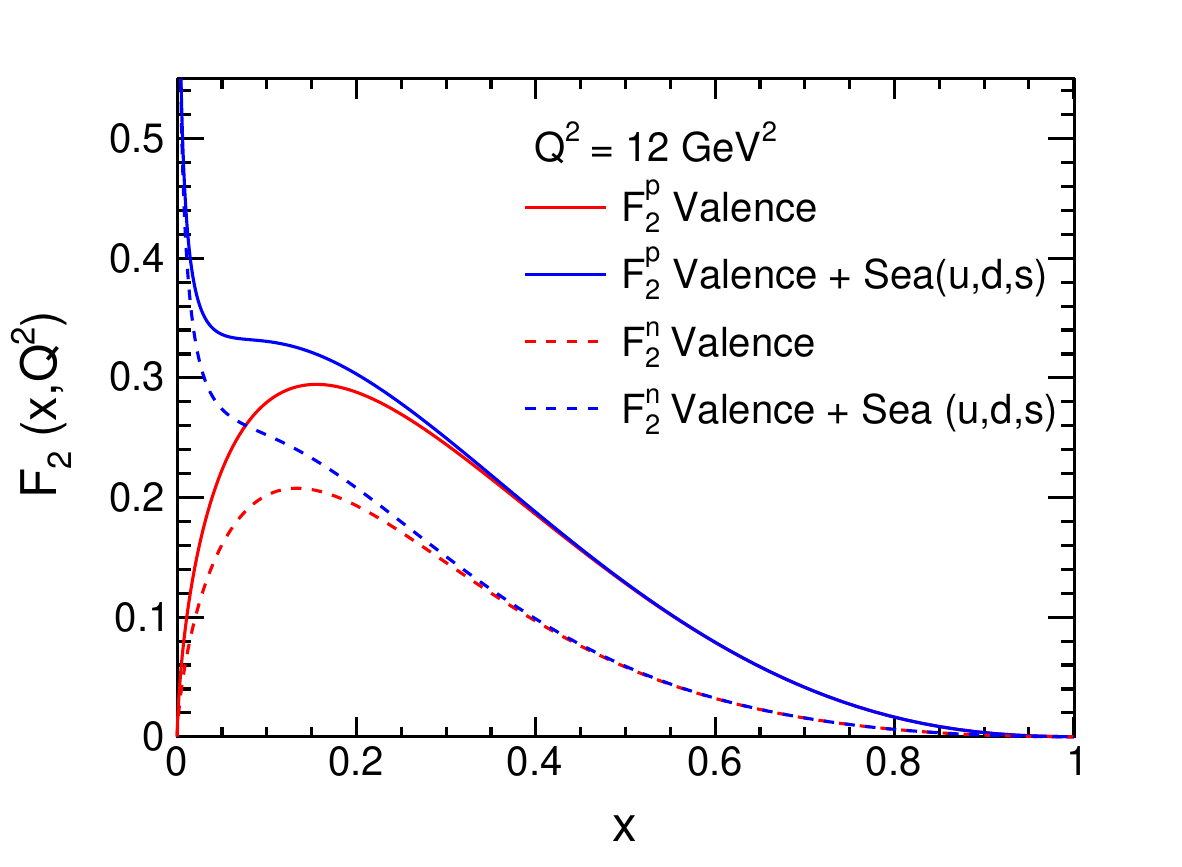}
    \caption{
      Comparisons of the structure functions of proton (solid curve) and neutron (dashed curve).
      The structure function calculations are performed with and without the sea quark distribution.
    }
    \label{F2p&F2n}
  \end{center}
\end{figure}

Figure~\ref{F2p&F2n} shows the comparisons between the obtained proton structure function
and neutron structure function, from MEM, at $Q^{2}$ = 12 GeV$^{2}$.
We see that the sea quark distributions only contribute significantly in the small $x$ region.

\subsection{Structure function ratio \( F_2^{n}/F_2^{p} \)}
\begin{figure}[htp]
\begin{center}
\includegraphics[width=0.45\textwidth]{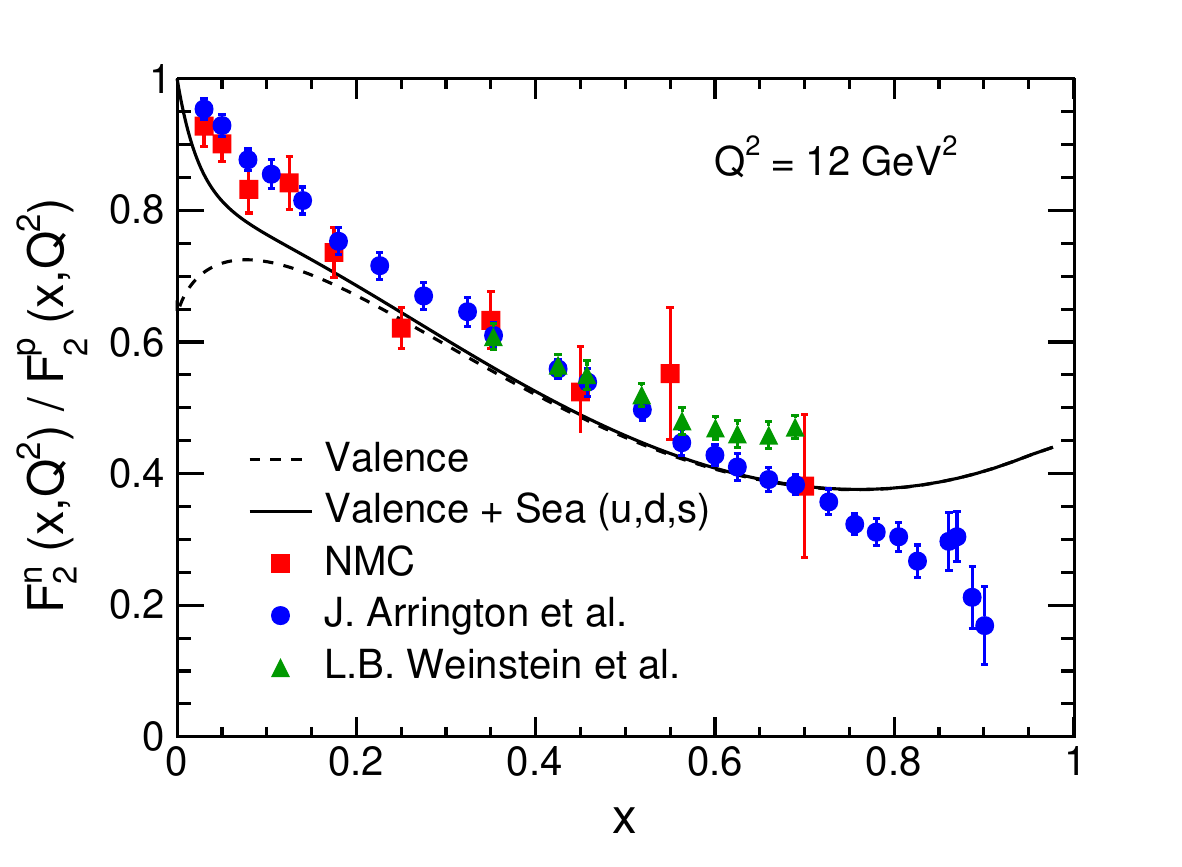}
\caption{
  The obtained structure function ratio of neutron to proton $F_{2}^{n}/F_{2}^{p}$
  from MEM, compared with the previous experimental extractions.
  NMC data (squares) are extracted from the simultaneous measurements on hydrogen and deuterium
  with the incident muon beam \cite{NewMuon:1991exl}.
  The analysis by J. Arrington et al. (circles) are corrected with the nucleon motion in deuterium \cite{Arrington:2008zh}.
  The triangles show $F^{n}_{2}/F^{p}_{2}$ data extracted from the deuteron in-medium correction \cite{Weinstein:2010rt}.
  The errors plotted display the total experimental uncertainties.
}
\label{F2n/F2p_old_exp._data}
\end{center}
\end{figure}

Figure \ref{F2n/F2p_old_exp._data} illustrates the predicted ratio \( F_2^{n}/F_2^{p} \) derived from the MEM as a function of the Bjorken variable \( x \),
juxtaposed with previous measurements obtained from deep inelastic scattering on deuteron and proton targets. The data from the New Muon Collaboration (NMC),
represented by squares, originate from DIS measurements employing a muon beam directed at hydrogen and deuterium \cite{NewMuon:1991exl}.
Additionally, the measurements by J. Arrington et al. (circles) provide a systematic analysis of the neutron structure function based on deuteron and proton data,
incorporating the impulse approximation and an effective two-nucleon mass operator \cite{Arrington:2008zh}. The triangle markers indicate the \( F_2^{n}/F_2^{p} \) ratio
extracted after applying deuteron in-medium corrections \cite{Weinstein:2010rt}.
In this figure, our predicted structure function ratio \( F_2^{n}/F_2^{p} \) from MEM is observed to be slightly lower than the corresponding experimental data,
yet remains consistent with global DIS measurements, except for the region where \( x > 0.7 \) as reported in Arrington's work. Notably,
Arrington's findings reveal a decreasing trend in the \( F_2^{n}/F_2^{p} \) ratio as \( x \) increases, a behavior that contrasts with our predictions.

The extraction of free neutron structure information from deuteron data necessitates consideration of various models for nuclear corrections.
These include the nuclear density-dependent EMC effect \cite{Whitlow:1991uw, Frankfurt:1988nt}, Fermi motion \cite{Whitlow:1991uw}, on-shell model extractions \cite{Gomez:1993ri},
and off-shell corrections \cite{Melnitchouk:1994rv,Melnitchouk:1995fc}. A significant challenge posed by these corrections is the inherent fact that deuterons and \(^3\)He, despite their weak binding,
do not accurately represent free neutron-proton systems. Furthermore, the employment of differing nuclear correction models can lead to substantial uncertainties
in the ratio \( F_2^{n}/F_2^{p} \) as \( x \) approaches 1, highlighting the complexities involved in this extraction process.

\begin{figure}[htp]
  \begin{center}
    \includegraphics[width=0.45\textwidth]{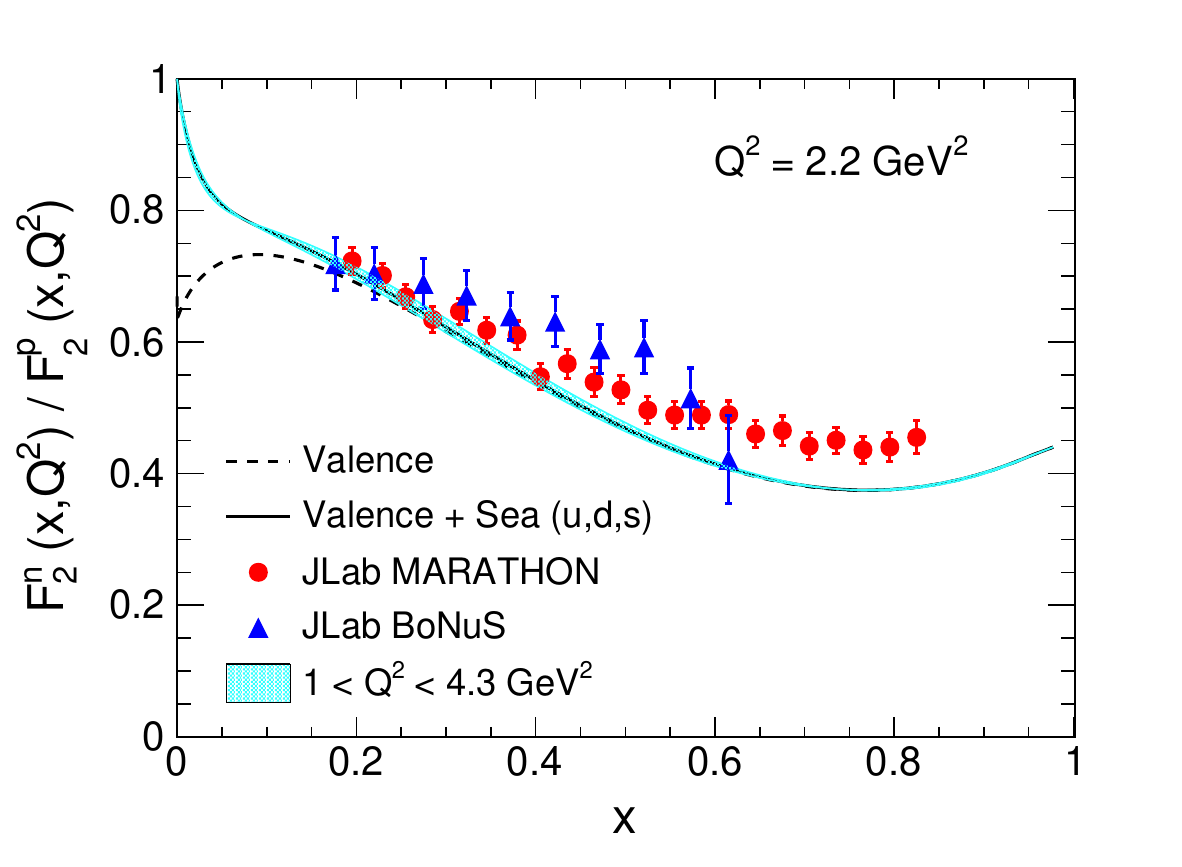}
    \caption{
      Comparisons of our predicted $F^{n}_{2}/F^{p}_{2}$ ratio
      with the JLab MARATHON experiment \cite{JeffersonLabHallATritium:2021usd} and BONuS experiment \cite{CLAS:2014jvt}.
      The error bars show the statistical, point to point systematic, and normalization uncertainties of the data.
      Our theoretical prediction is at $Q^{2} = 2.2$ GeV$^{2}$, and the cyan band shows the variation
      of the scale evolution from 1 GeV$^2$ to 4.3 GeV$^2$ (in accordance with the experimental data).
    }
    \label{F2n/F2p_MARATHON_BONuS}
  \end{center}
\end{figure}

Figure~\ref{F2n/F2p_MARATHON_BONuS} shows the predicted ratio $F^{n}_{2}/F^{p}_{2}$ at $Q^{2} = 2.2$ GeV$^{2}$,
compared with measurements from the JLab MARATHON experiment \cite{JeffersonLabHallATritium:2021usd} and the BONuS experiment \cite{CLAS:2014jvt}.
The cyan band of the prediction shows the variation of the $Q^2$-dependence from 1 GeV$^2$ to 4.3 GeV$^2$.
With the quark-hadron duality constrain under the \textcolor{red}{assumption of the} ``scaling" of the elastic form factors
as $q^{2} \rightarrow \infty$, our prediction of the neutron and proton structure function ratio $F^{n}_{2}/F^{p}_{2}$  is in excellent agreement with the experimental data.
From the Fig.~\ref{F2n/F2p_MARATHON_BONuS}, we can see the obvious hadron resonance peaks in BONuS data in the $0.4 < x < 0.6$ region,
%even under the invariant mass of the final hadronic state cut of $W^{*} > 1.8$ GeV/$c^{2}$ \cite{CLAS:2014jvt}.
\textcolor{red}{even with the cut $W^{*}> 1.8$ GeV/$c^{2}$ \cite{CLAS:2014jvt}, where $W^{*}$ here denotes the invariant mass of the final hadrons.}
This is due to the significant non-perturbative quark-gluon interaction and the inclusive lepton-nucleon cross section dominated by nucleon resonance at lower energies \cite{Niculescu:2015wka}.
However, the JLab MARATHON experiment from measurements of deep inelastic scattering of electrons from $^{3}$H and $^{3}$He nuclei eliminates this discrepancy.
In Fig.~\ref{F2n/F2p_MARATHON_BONuS}, by comparison,
we can find that there is a slight difference in magnitude between our \textcolor{red}{prediction} and the JLab MARATHON when $x>0.5$.
Our predicted $F^{n}_{2}/F^{p}_{2}$ ratio \textcolor{red}{follows} the same distribution trend as JLab MARATHON measurements,
and the ratio of neutron to proton structure functions stops decreasing with $x$ increases.
In short, the MEM prediction of the ratio $F^{n}_{2}/F^{p}_{2}$ successfully describes the measurements of JLab MARATHON and JLab BONuS
if the quark-hadron duality assumption is taken into account \cite{Bloom:1971ye,Niculescu:2015wka}, especially in the region of $x > 0.7$.

\subsection{$u/d$ ratio of neutron}
\begin{figure}[htp]
\begin{center}
\includegraphics[width=0.45\textwidth]{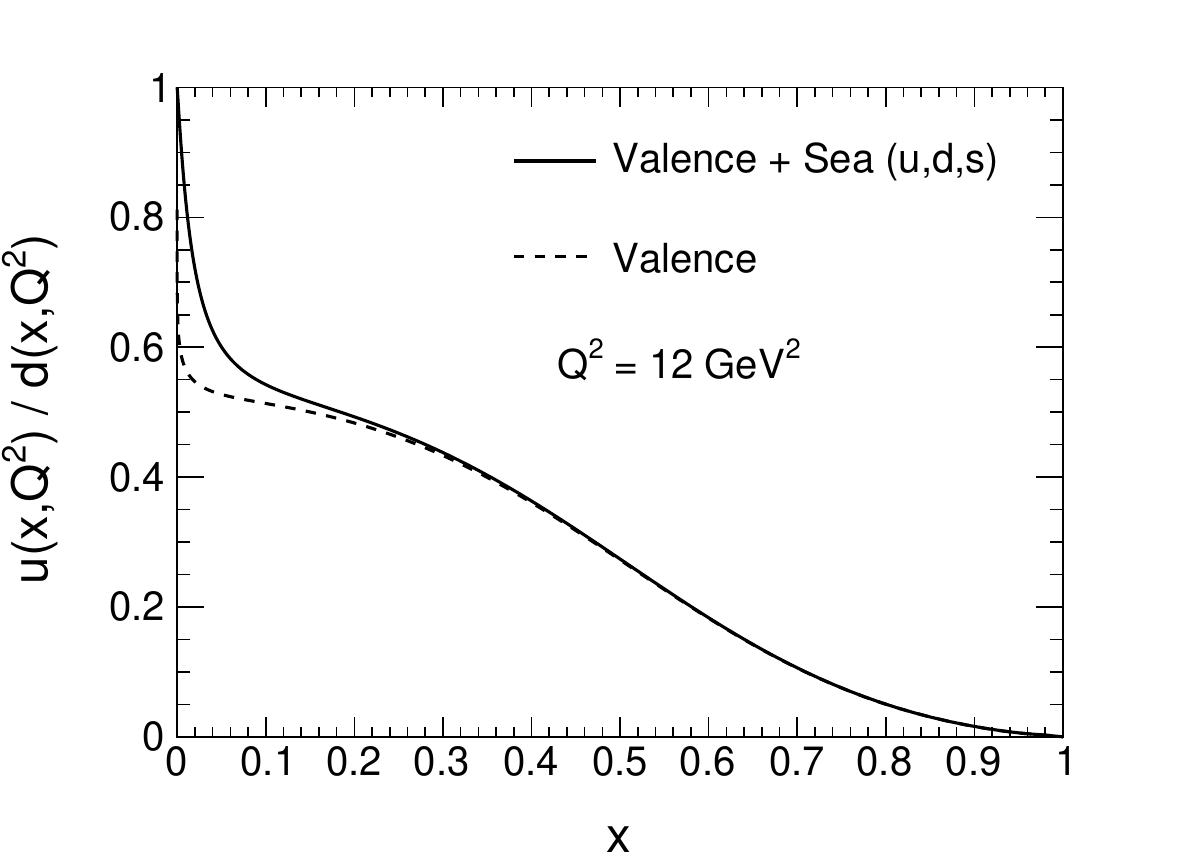}
\caption{
  The $u/d$ ratio of the neutron at $Q^{2} = 12$ GeV$^{2}$ given by MEM.
  The solid curve includes also the sea quark contributions,
  while the dashed curve is calculated from the valence quark distributions only. 
}
\label{n_u/d}
\end{center}
\end{figure}

The $u/d$ ratio is a quantity which is directly related to the ratio $F^{n}_{2}/F^{p}_{2}$.
Figure~\ref{n_u/d} shows the predicted $u/d$ ratio as a function of $x$ at $Q^{2} = 12$ GeV$^{2}$.
MEM in this work gives $u/d=0$ at the limit of $x$ approaching one.
In experiment, the $d/u$ ratio of the proton is usually extracted from the $F^{n}_{2}/F^{p}_{2}$ data,
neglecting the strange quark contribution.
Note that the $d/u$ ratio of the proton equals the $u/d$ ratio of the neutron
under the assumption of isospin symmetry.

Table \ref{F2n/F2p&u/d_ratio} lists the comparisons of our predicted $F_2^{n}/F_{2}^{p}$ ratio
and $u/d$ ratio of neutron at the limit of $x \rightarrow 1$, compared with other theoretical predictions.
The structure function ratio between neutron and proton is denoted as
$R^{n/p}(x) = {F^{n}_2}(x)/{F_2^{p}}(x)$ in this work for the convenience of discussions.
In this work by MEM, we find that $R^{n/p}(0) \approx 1.0$ \textcolor{red}{and} $R^{n/p}(1) \approx 0.47$,
which \textcolor{red}{are} clearly depicted in Fig.~\ref{F2n/F2p_old_exp._data}.
$R^{n/p}(0) \approx 1.0$ is due to the fact that the sea quark distribution dominates at small $x$.
The other result is that $u^{n}/d^{n}\approx 0$ when $x$ approaches one.
Our result is quite different from the predictions from the exact spin-flavor SU(6) symmetry,
the perturbative QCD assuming the helicity conserved quarks interacted via hard gluon exchange \cite{Farrar:1975yb},
and the quark counting rule method \cite{Brodsky:1994kg}.
However the models with the SU(6) symmetry breaking via the scalar diquark dominance \cite{feynman2018photon, Close:1973xw}
predict the very similar result of ours.
In addition, the hyperfine-perturbed quark model also makes a prediction that similar to ours \cite{Isgur:1998yb}.  
These predictions at large $x$ are needed to be tested in the future experiments.

\begin{table}
  \caption{The list of the theoretical predictions of $R^{n/p}=F_2^{n}/F_{2}^{p}$
           and $u^{n}/d^{n}$ at the limit of $x\rightarrow 1$ from different models.
           }
  \label{Tab:ccs}
  \begin{tabular}{ccccccccccccccccccc}
    \hline\hline
    &&&  $R^{n/p}$             &&&&&    $u^{n}/d^{n}$                      \\ \hline
    SU(6) Flavor Symmetry&&&  2/3                  &&&&&    1/2     &                   \\
    Perturbative QCD     &&&  3/7                  &&&&&    1/5     &                   \\
    Quark Counting Rule  &&&  3/7                  &&&&&    1/5     &                   \\
    Diquark (Feynman)    &&&  1/4                  &&&&&    0       &                   \\
    Quark Model (Isgur)  &&&  1/4                  &&&&&    0       &                   \\
    MEM (this work)      &&&  0.47                 &&&&&    0       &                   \\
    \hline\hline
    \label{F2n/F2p&u/d_ratio}
  \end{tabular}
\end{table}

\subsection{Assessment of isospin symmetry}
\begin{figure}[htp]
  \begin{center}
    \includegraphics[width=0.45\textwidth]{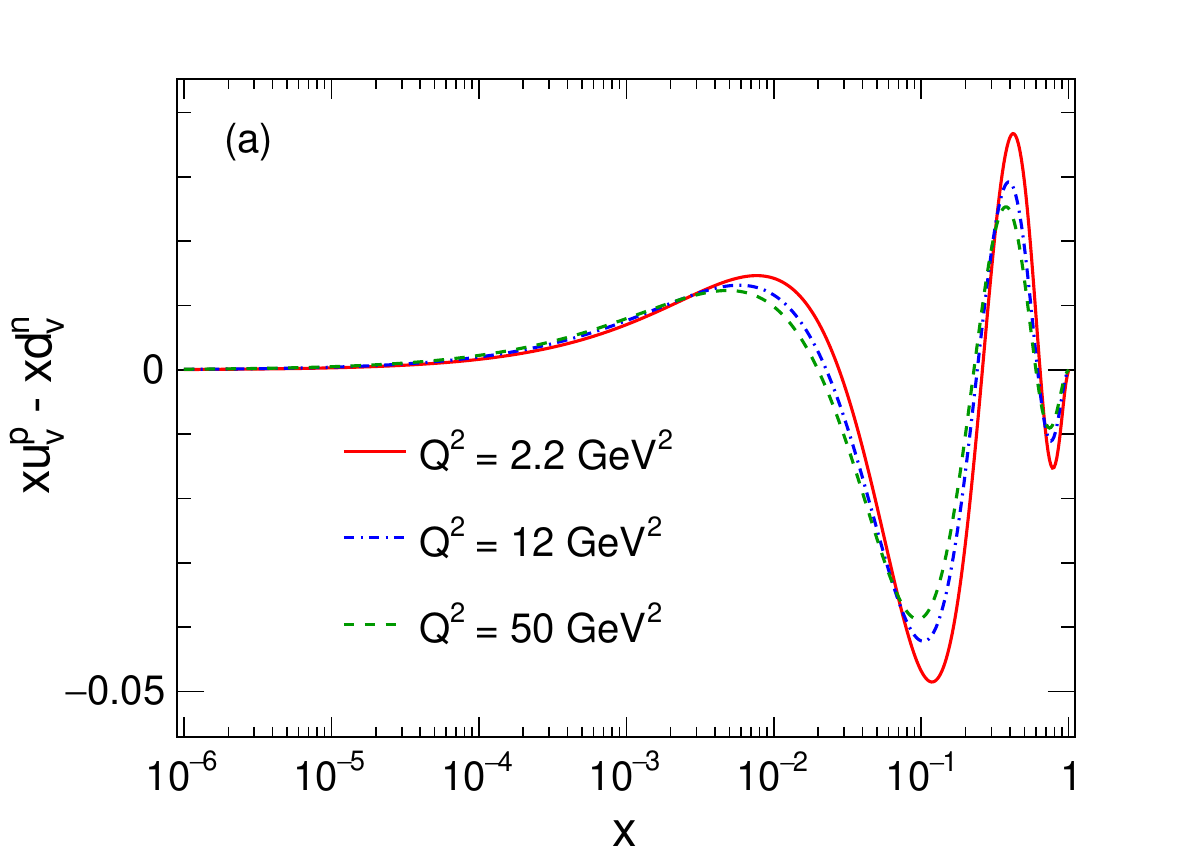}
    \includegraphics[width=0.46\textwidth]{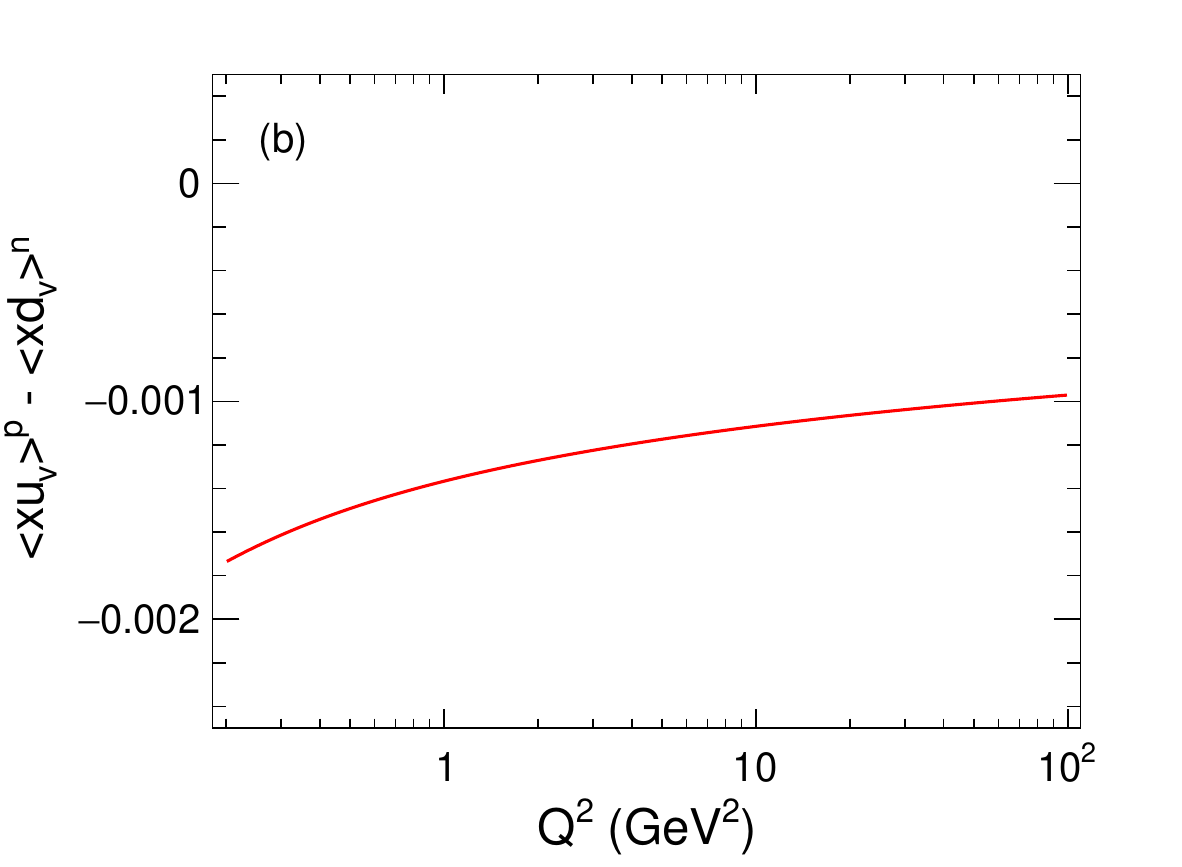}
    \caption{
      Panel (a) The valence quark distribution difference between proton and neutron as a function of Bjorken variable $x$.
      Panel (b) The difference of the first-order moments of valence quark distributions between proton and neutron as a function of $Q^{2}$.
    }
    \label{DvOverUv}
  \end{center}
\end{figure}

To investigate the degree of isospin symmetry between the proton and neutron, we have examined the differences in their valence quark distributions and first-order moments. These comparisons are presented in Fig.~\ref{DvOverUv}, where panel (a) and panel (b) display the differences in valence quark distributions and first-order moments, respectively. By analyzing the results, it is evident that at intermediate values of $x$ (0.02 $ < x < $ 0.5), the isospin symmetry between the proton and neutron is indeed violated, whereas at the limits $x \sim 0$ and $x = 1$, the symmetry appears to be preserved.

Furthermore, inspection of the first-order moments shown in panel (b) reveals a maximal difference of 0.0017 between the two nucleons at low $Q^2$, and a gradual recovery of the isospin symmetry as $Q^2$ increases. While the results in Fig.~\ref{DvOverUv} clearly demonstrate the presence of isospin breaking, the magnitude of this effect is found to be moderate.

The minor isospin symmetry breaking between the proton and neutron could be attributed to differences in their initial valence quark distributions, as well as variations in their charge radii and masses. These differences in nucleon properties may contribute to the observed isospin breaking and warrant further investigation to clarify their role in the quark structure of the nucleon.

\section{Summary}
\label{summary}

The maximum entropy method offers a novel and efficient approach for predicting the nonperturbative structure of nucleons, particularly for neutron,
where experimental data are scarce. By imposing constraints from quark models, quark-hadron duality, and quark confinement,
MEM can be used to determine a reasonable initial valence quark distribution for the neutron. Our analysis reveals that at an initial scale $Q_0^2$,
the neutron is composed of three valence quarks with no sea quark or gluon distributions.
The predicted ratio of the valence structure functions between neutron and proton,
$F_2^{n}/F_{2}^{p}$, is found to be consistent with current JLab experimental data and world deep inelastic scattering data.
The nearly free neutron measurements by the JLab MARATHON and BONuS collaborations provide a crucial test of our predictions from MEM.
By considering the contamination from nucleon resonances, our results show that the obtained $F_2^{n}/F_{2}^{p}$ ratio is in agreement with JLab BONuS experimental results,
while the higher energy MARATHON measurements remove the contamination and provide a clear test of our predictions.
It is worth noting that the $F_2^{n}/F_{2}^{p}$ ratio obtained based on the quark-hadron duality assumption
is consistent with the JLab MARATHON experimental measurements in the large Bjoken variable $x$ region.
Furthermore, our analysis reveals that
the large-$x$ behaviors of $F_2^{n}/F_{2}^{p}$ and $u^{n}/d^{n}$ are consistent with perturbative QCD and quark counting rule predictions, but deviate from SU(6) flavor symmetry,
diquark model, and quark model predictions. Finally, our results show that the isospin breaking between proton and neutron is minor, and may arise from differences in initial valence quark input,
charge radius, and mass between proton and neutron.

\begin{acknowledgments}
This work is supported by the National Natural Science Foundation of China under the Grant NO. 12305127,
the International Partnership Program of the Chinese Academy of Sciences under the Grant NO. 016GJHZ2022054FN.
and National Key R$\&$D Program of China under the Grant NO. 2024YFE0109802.
\end{acknowledgments}

\section*{Data Availability Statement}
Data will be made available on reasonable request.

\bibliographystyle{apsrev4-1}
\bibliography{refs.bib}

\begin{thebibliography}{47}%
\makeatletter
\providecommand \@ifxundefined [1]{%
 \@ifx{#1\undefined}
}%
\providecommand \@ifnum [1]{%
 \ifnum #1\expandafter \@firstoftwo
 \else \expandafter \@secondoftwo
 \fi
}%
\providecommand \@ifx [1]{%
 \ifx #1\expandafter \@firstoftwo
 \else \expandafter \@secondoftwo
 \fi
}%
\providecommand \natexlab [1]{#1}%
\providecommand \enquote  [1]{``#1''}%
\providecommand \bibnamefont  [1]{#1}%
\providecommand \bibfnamefont [1]{#1}%
\providecommand \citenamefont [1]{#1}%
\providecommand \href@noop [0]{\@secondoftwo}%
\providecommand \href [0]{\begingroup \@sanitize@url \@href}%
\providecommand \@href[1]{\@@startlink{#1}\@@href}%
\providecommand \@@href[1]{\endgroup#1\@@endlink}%
\providecommand \@sanitize@url [0]{\catcode `\\12\catcode `\$12\catcode
  `\&12\catcode `\#12\catcode `\^12\catcode `\_12\catcode `\%12\relax}%
\providecommand \@@startlink[1]{}%
\providecommand \@@endlink[0]{}%
\providecommand \url  [0]{\begingroup\@sanitize@url \@url }%
\providecommand \@url [1]{\endgroup\@href {#1}{\urlprefix }}%
\providecommand \urlprefix  [0]{URL }%
\providecommand \Eprint [0]{\href }%
\providecommand \doibase [0]{http://dx.doi.org/}%
\providecommand \selectlanguage [0]{\@gobble}%
\providecommand \bibinfo  [0]{\@secondoftwo}%
\providecommand \bibfield  [0]{\@secondoftwo}%
\providecommand \translation [1]{[#1]}%
\providecommand \BibitemOpen [0]{}%
\providecommand \bibitemStop [0]{}%
\providecommand \bibitemNoStop [0]{.\EOS\space}%
\providecommand \EOS [0]{\spacefactor3000\relax}%
\providecommand \BibitemShut  [1]{\csname bibitem#1\endcsname}%
\let\auto@bib@innerbib\@empty
%</preamble>
\bibitem [{\citenamefont {Amaudruz}\ \emph {et~al.}(1992)\citenamefont
  {Amaudruz} \emph {et~al.}}]{NewMuon:1991exl}%
  \BibitemOpen
  \bibfield  {author} {\bibinfo {author} {\bibfnamefont {P.}~\bibnamefont
  {Amaudruz}} \emph {et~al.} (\bibinfo {collaboration} {New Muon}),\ }\href
  {\doibase 10.1016/0550-3213(92)90227-3} {\bibfield  {journal} {\bibinfo
  {journal} {Nucl. Phys. B}\ }\textbf {\bibinfo {volume} {371}},\ \bibinfo
  {pages} {3} (\bibinfo {year} {1992})}\BibitemShut {NoStop}%
\bibitem [{\citenamefont {Arrington}\ \emph {et~al.}(2009)\citenamefont
  {Arrington}, \citenamefont {Coester}, \citenamefont {Holt},\ and\
  \citenamefont {Lee}}]{Arrington:2008zh}%
  \BibitemOpen
  \bibfield  {author} {\bibinfo {author} {\bibfnamefont {J.}~\bibnamefont
  {Arrington}}, \bibinfo {author} {\bibfnamefont {F.}~\bibnamefont {Coester}},
  \bibinfo {author} {\bibfnamefont {R.~J.}\ \bibnamefont {Holt}}, \ and\
  \bibinfo {author} {\bibfnamefont {T.~S.~H.}\ \bibnamefont {Lee}},\ }\href
  {\doibase 10.1088/0954-3899/36/2/025005} {\bibfield  {journal} {\bibinfo
  {journal} {J. Phys. G}\ }\textbf {\bibinfo {volume} {36}},\ \bibinfo {pages}
  {025005} (\bibinfo {year} {2009})},\ \Eprint {http://arxiv.org/abs/0805.3116}
  {arXiv:0805.3116 [nucl-th]} \BibitemShut {NoStop}%
\bibitem [{\citenamefont {Weinstein}\ \emph {et~al.}(2011)\citenamefont
  {Weinstein}, \citenamefont {Piasetzky}, \citenamefont {Higinbotham},
  \citenamefont {Gomez}, \citenamefont {Hen},\ and\ \citenamefont
  {Shneor}}]{Weinstein:2010rt}%
  \BibitemOpen
  \bibfield  {author} {\bibinfo {author} {\bibfnamefont {L.~B.}\ \bibnamefont
  {Weinstein}}, \bibinfo {author} {\bibfnamefont {E.}~\bibnamefont
  {Piasetzky}}, \bibinfo {author} {\bibfnamefont {D.~W.}\ \bibnamefont
  {Higinbotham}}, \bibinfo {author} {\bibfnamefont {J.}~\bibnamefont {Gomez}},
  \bibinfo {author} {\bibfnamefont {O.}~\bibnamefont {Hen}}, \ and\ \bibinfo
  {author} {\bibfnamefont {R.}~\bibnamefont {Shneor}},\ }\href {\doibase
  10.1103/PhysRevLett.106.052301} {\bibfield  {journal} {\bibinfo  {journal}
  {Phys. Rev. Lett.}\ }\textbf {\bibinfo {volume} {106}},\ \bibinfo {pages}
  {052301} (\bibinfo {year} {2011})},\ \Eprint {http://arxiv.org/abs/1009.5666}
  {arXiv:1009.5666 [hep-ph]} \BibitemShut {NoStop}%
\bibitem [{\citenamefont {Arrington}\ \emph {et~al.}(2012)\citenamefont
  {Arrington}, \citenamefont {Rubin},\ and\ \citenamefont
  {Melnitchouk}}]{Arrington:2011qt}%
  \BibitemOpen
  \bibfield  {author} {\bibinfo {author} {\bibfnamefont {J.}~\bibnamefont
  {Arrington}}, \bibinfo {author} {\bibfnamefont {J.~G.}\ \bibnamefont
  {Rubin}}, \ and\ \bibinfo {author} {\bibfnamefont {W.}~\bibnamefont
  {Melnitchouk}},\ }\href {\doibase 10.1103/PhysRevLett.108.252001} {\bibfield
  {journal} {\bibinfo  {journal} {Phys. Rev. Lett.}\ }\textbf {\bibinfo
  {volume} {108}},\ \bibinfo {pages} {252001} (\bibinfo {year} {2012})},\
  \Eprint {http://arxiv.org/abs/1110.3362} {arXiv:1110.3362 [hep-ph]}
  \BibitemShut {NoStop}%
\bibitem [{\citenamefont {Baillie}\ \emph {et~al.}(2012)\citenamefont {Baillie}
  \emph {et~al.}}]{CLAS:2011qvj}%
  \BibitemOpen
  \bibfield  {author} {\bibinfo {author} {\bibfnamefont {N.}~\bibnamefont
  {Baillie}} \emph {et~al.} (\bibinfo {collaboration} {CLAS}),\ }\href
  {\doibase 10.1103/PhysRevLett.108.142001} {\bibfield  {journal} {\bibinfo
  {journal} {Phys. Rev. Lett.}\ }\textbf {\bibinfo {volume} {108}},\ \bibinfo
  {pages} {142001} (\bibinfo {year} {2012})},\ \bibinfo {note} {[Erratum:
  Phys.Rev.Lett. 108, 199902 (2012)]},\ \Eprint
  {http://arxiv.org/abs/1110.2770} {arXiv:1110.2770 [nucl-ex]} \BibitemShut
  {NoStop}%
\bibitem [{\citenamefont {Hen}\ \emph {et~al.}(2011)\citenamefont {Hen},
  \citenamefont {Piasetzky}, \citenamefont {Shneor}, \citenamefont
  {Weinstein},\ and\ \citenamefont {Higinbotham}}]{Hen:2011ad}%
  \BibitemOpen
  \bibfield  {author} {\bibinfo {author} {\bibfnamefont {O.}~\bibnamefont
  {Hen}}, \bibinfo {author} {\bibfnamefont {E.}~\bibnamefont {Piasetzky}},
  \bibinfo {author} {\bibfnamefont {R.}~\bibnamefont {Shneor}}, \bibinfo
  {author} {\bibfnamefont {L.~B.}\ \bibnamefont {Weinstein}}, \ and\ \bibinfo
  {author} {\bibfnamefont {D.~W.}\ \bibnamefont {Higinbotham}},\ }\href@noop {}
  {\  (\bibinfo {year} {2011})},\ \Eprint {http://arxiv.org/abs/1109.6197}
  {arXiv:1109.6197 [hep-ph]} \BibitemShut {NoStop}%
\bibitem [{\citenamefont {Tkachenko}\ \emph {et~al.}(2014)\citenamefont
  {Tkachenko} \emph {et~al.}}]{CLAS:2014jvt}%
  \BibitemOpen
  \bibfield  {author} {\bibinfo {author} {\bibfnamefont {S.}~\bibnamefont
  {Tkachenko}} \emph {et~al.} (\bibinfo {collaboration} {CLAS}),\ }\href
  {\doibase 10.1103/PhysRevC.89.045206} {\bibfield  {journal} {\bibinfo
  {journal} {Phys. Rev. C}\ }\textbf {\bibinfo {volume} {89}},\ \bibinfo
  {pages} {045206} (\bibinfo {year} {2014})},\ \bibinfo {note} {[Addendum:
  Phys.Rev.C 90, 059901 (2014)]},\ \Eprint {http://arxiv.org/abs/1402.2477}
  {arXiv:1402.2477 [nucl-ex]} \BibitemShut {NoStop}%
\bibitem [{\citenamefont {Niculescu}\ \emph {et~al.}(2015)\citenamefont
  {Niculescu} \emph {et~al.}}]{Niculescu:2015wka}%
  \BibitemOpen
  \bibfield  {author} {\bibinfo {author} {\bibfnamefont {I.}~\bibnamefont
  {Niculescu}} \emph {et~al.},\ }\href {\doibase 10.1103/PhysRevC.91.055206}
  {\bibfield  {journal} {\bibinfo  {journal} {Phys. Rev. C}\ }\textbf {\bibinfo
  {volume} {91}},\ \bibinfo {pages} {055206} (\bibinfo {year} {2015})},\
  \Eprint {http://arxiv.org/abs/1501.02203} {arXiv:1501.02203 [hep-ex]}
  \BibitemShut {NoStop}%
\bibitem [{\citenamefont {Accardi}\ \emph {et~al.}(2016)\citenamefont
  {Accardi}, \citenamefont {Brady}, \citenamefont {Melnitchouk}, \citenamefont
  {Owens},\ and\ \citenamefont {Sato}}]{Accardi:2016qay}%
  \BibitemOpen
  \bibfield  {author} {\bibinfo {author} {\bibfnamefont {A.}~\bibnamefont
  {Accardi}}, \bibinfo {author} {\bibfnamefont {L.~T.}\ \bibnamefont {Brady}},
  \bibinfo {author} {\bibfnamefont {W.}~\bibnamefont {Melnitchouk}}, \bibinfo
  {author} {\bibfnamefont {J.~F.}\ \bibnamefont {Owens}}, \ and\ \bibinfo
  {author} {\bibfnamefont {N.}~\bibnamefont {Sato}},\ }\href {\doibase
  10.1103/PhysRevD.93.114017} {\bibfield  {journal} {\bibinfo  {journal} {Phys.
  Rev. D}\ }\textbf {\bibinfo {volume} {93}},\ \bibinfo {pages} {114017}
  (\bibinfo {year} {2016})},\ \Eprint {http://arxiv.org/abs/1602.03154}
  {arXiv:1602.03154 [hep-ph]} \BibitemShut {NoStop}%
\bibitem [{\citenamefont {Szumila-Vance}\ \emph {et~al.}(2021)\citenamefont
  {Szumila-Vance}, \citenamefont {Keppel}, \citenamefont {Escalante},\ and\
  \citenamefont {Kalantarians}}]{Szumila-Vance:2020zpt}%
  \BibitemOpen
  \bibfield  {author} {\bibinfo {author} {\bibfnamefont {H.}~\bibnamefont
  {Szumila-Vance}}, \bibinfo {author} {\bibfnamefont {C.}~\bibnamefont
  {Keppel}}, \bibinfo {author} {\bibfnamefont {S.}~\bibnamefont {Escalante}}, \
  and\ \bibinfo {author} {\bibfnamefont {N.}~\bibnamefont {Kalantarians}},\
  }\href {\doibase 10.1103/PhysRevC.103.015201} {\bibfield  {journal} {\bibinfo
   {journal} {Phys. Rev. C}\ }\textbf {\bibinfo {volume} {103}},\ \bibinfo
  {pages} {015201} (\bibinfo {year} {2021})},\ \Eprint
  {http://arxiv.org/abs/2002.02597} {arXiv:2002.02597 [nucl-ex]} \BibitemShut
  {NoStop}%
\bibitem [{\citenamefont {Abrams}\ \emph {et~al.}(2022)\citenamefont {Abrams}
  \emph {et~al.}}]{JeffersonLabHallATritium:2021usd}%
  \BibitemOpen
  \bibfield  {author} {\bibinfo {author} {\bibfnamefont {D.}~\bibnamefont
  {Abrams}} \emph {et~al.} (\bibinfo {collaboration} {Jefferson Lab Hall A
  Tritium}),\ }\href {\doibase 10.1103/PhysRevLett.128.132003} {\bibfield
  {journal} {\bibinfo  {journal} {Phys. Rev. Lett.}\ }\textbf {\bibinfo
  {volume} {128}},\ \bibinfo {pages} {132003} (\bibinfo {year} {2022})},\
  \Eprint {http://arxiv.org/abs/2104.05850} {arXiv:2104.05850 [hep-ex]}
  \BibitemShut {NoStop}%
\bibitem [{\citenamefont {Fenker}\ \emph {et~al.}(2008)\citenamefont {Fenker}
  \emph {et~al.}}]{Fenker:2008zz}%
  \BibitemOpen
  \bibfield  {author} {\bibinfo {author} {\bibfnamefont {H.~C.}\ \bibnamefont
  {Fenker}} \emph {et~al.},\ }\href {\doibase 10.1016/j.nima.2008.04.047}
  {\bibfield  {journal} {\bibinfo  {journal} {Nucl. Instrum. Meth. A}\ }\textbf
  {\bibinfo {volume} {592}},\ \bibinfo {pages} {273} (\bibinfo {year}
  {2008})}\BibitemShut {NoStop}%
\bibitem [{\citenamefont {Li}\ \emph {et~al.}(2024)\citenamefont {Li},
  \citenamefont {Accardi}, \citenamefont {Cerutti}, \citenamefont {Fernando},
  \citenamefont {Keppel}, \citenamefont {Melnitchouk}, \citenamefont
  {Monaghan}, \citenamefont {Niculescu}, \citenamefont {Niculescu},\ and\
  \citenamefont {Owens}}]{Li:2023yda}%
  \BibitemOpen
  \bibfield  {author} {\bibinfo {author} {\bibfnamefont {S.}~\bibnamefont
  {Li}}, \bibinfo {author} {\bibfnamefont {A.}~\bibnamefont {Accardi}},
  \bibinfo {author} {\bibfnamefont {M.}~\bibnamefont {Cerutti}}, \bibinfo
  {author} {\bibfnamefont {I.~P.}\ \bibnamefont {Fernando}}, \bibinfo {author}
  {\bibfnamefont {C.~E.}\ \bibnamefont {Keppel}}, \bibinfo {author}
  {\bibfnamefont {W.}~\bibnamefont {Melnitchouk}}, \bibinfo {author}
  {\bibfnamefont {P.}~\bibnamefont {Monaghan}}, \bibinfo {author}
  {\bibfnamefont {G.}~\bibnamefont {Niculescu}}, \bibinfo {author}
  {\bibfnamefont {M.~I.}\ \bibnamefont {Niculescu}}, \ and\ \bibinfo {author}
  {\bibfnamefont {J.~F.}\ \bibnamefont {Owens}},\ }\href {\doibase
  10.1103/PhysRevD.109.074036} {\bibfield  {journal} {\bibinfo  {journal}
  {Phys. Rev. D}\ }\textbf {\bibinfo {volume} {109}},\ \bibinfo {pages}
  {074036} (\bibinfo {year} {2024})},\ \Eprint
  {http://arxiv.org/abs/2309.16851} {arXiv:2309.16851 [hep-ph]} \BibitemShut
  {NoStop}%
\bibitem [{\citenamefont {Whitlow}\ \emph {et~al.}(1992)\citenamefont
  {Whitlow}, \citenamefont {Riordan}, \citenamefont {Dasu}, \citenamefont
  {Rock},\ and\ \citenamefont {Bodek}}]{Whitlow:1991uw}%
  \BibitemOpen
  \bibfield  {author} {\bibinfo {author} {\bibfnamefont {L.~W.}\ \bibnamefont
  {Whitlow}}, \bibinfo {author} {\bibfnamefont {E.~M.}\ \bibnamefont
  {Riordan}}, \bibinfo {author} {\bibfnamefont {S.}~\bibnamefont {Dasu}},
  \bibinfo {author} {\bibfnamefont {S.}~\bibnamefont {Rock}}, \ and\ \bibinfo
  {author} {\bibfnamefont {A.}~\bibnamefont {Bodek}},\ }\href {\doibase
  10.1016/0370-2693(92)90672-Q} {\bibfield  {journal} {\bibinfo  {journal}
  {Phys. Lett. B}\ }\textbf {\bibinfo {volume} {282}},\ \bibinfo {pages} {475}
  (\bibinfo {year} {1992})}\BibitemShut {NoStop}%
\bibitem [{\citenamefont {Frankfurt}\ and\ \citenamefont
  {Strikman}(1988)}]{Frankfurt:1988nt}%
  \BibitemOpen
  \bibfield  {author} {\bibinfo {author} {\bibfnamefont {L.~L.}\ \bibnamefont
  {Frankfurt}}\ and\ \bibinfo {author} {\bibfnamefont {M.~I.}\ \bibnamefont
  {Strikman}},\ }\href {\doibase 10.1016/0370-1573(88)90179-2} {\bibfield
  {journal} {\bibinfo  {journal} {Phys. Rept.}\ }\textbf {\bibinfo {volume}
  {160}},\ \bibinfo {pages} {235} (\bibinfo {year} {1988})}\BibitemShut
  {NoStop}%
\bibitem [{\citenamefont {Gomez}\ \emph {et~al.}(1994)\citenamefont {Gomez}
  \emph {et~al.}}]{Gomez:1993ri}%
  \BibitemOpen
  \bibfield  {author} {\bibinfo {author} {\bibfnamefont {J.}~\bibnamefont
  {Gomez}} \emph {et~al.},\ }\href {\doibase 10.1103/PhysRevD.49.4348}
  {\bibfield  {journal} {\bibinfo  {journal} {Phys. Rev. D}\ }\textbf {\bibinfo
  {volume} {49}},\ \bibinfo {pages} {4348} (\bibinfo {year}
  {1994})}\BibitemShut {NoStop}%
\bibitem [{\citenamefont {Melnitchouk}\ \emph {et~al.}(1994)\citenamefont
  {Melnitchouk}, \citenamefont {Schreiber},\ and\ \citenamefont
  {Thomas}}]{Melnitchouk:1994rv}%
  \BibitemOpen
  \bibfield  {author} {\bibinfo {author} {\bibfnamefont {W.}~\bibnamefont
  {Melnitchouk}}, \bibinfo {author} {\bibfnamefont {A.~W.}\ \bibnamefont
  {Schreiber}}, \ and\ \bibinfo {author} {\bibfnamefont {A.~W.}\ \bibnamefont
  {Thomas}},\ }\href {\doibase 10.1016/0370-2693(94)91550-4} {\bibfield
  {journal} {\bibinfo  {journal} {Phys. Lett. B}\ }\textbf {\bibinfo {volume}
  {335}},\ \bibinfo {pages} {11} (\bibinfo {year} {1994})},\ \Eprint
  {http://arxiv.org/abs/nucl-th/9407007} {arXiv:nucl-th/9407007} \BibitemShut
  {NoStop}%
\bibitem [{\citenamefont {Melnitchouk}\ and\ \citenamefont
  {Thomas}(1996)}]{Melnitchouk:1995fc}%
  \BibitemOpen
  \bibfield  {author} {\bibinfo {author} {\bibfnamefont {W.}~\bibnamefont
  {Melnitchouk}}\ and\ \bibinfo {author} {\bibfnamefont {A.~W.}\ \bibnamefont
  {Thomas}},\ }\href {\doibase 10.1016/0370-2693(96)00292-4} {\bibfield
  {journal} {\bibinfo  {journal} {Phys. Lett. B}\ }\textbf {\bibinfo {volume}
  {377}},\ \bibinfo {pages} {11} (\bibinfo {year} {1996})},\ \Eprint
  {http://arxiv.org/abs/nucl-th/9602038} {arXiv:nucl-th/9602038} \BibitemShut
  {NoStop}%
\bibitem [{\citenamefont {Nakano}\ and\ \citenamefont
  {Wong}(1991)}]{Nakano:1991kh}%
  \BibitemOpen
  \bibfield  {author} {\bibinfo {author} {\bibfnamefont {K.}~\bibnamefont
  {Nakano}}\ and\ \bibinfo {author} {\bibfnamefont {S.~S.~M.}\ \bibnamefont
  {Wong}},\ }\href {\doibase 10.1016/0375-9474(91)90769-3} {\bibfield
  {journal} {\bibinfo  {journal} {Nucl. Phys. A}\ }\textbf {\bibinfo {volume}
  {530}},\ \bibinfo {pages} {555} (\bibinfo {year} {1991})}\BibitemShut
  {NoStop}%
\bibitem [{\citenamefont {Holt}\ and\ \citenamefont
  {Roberts}(2010)}]{Holt:2010vj}%
  \BibitemOpen
  \bibfield  {author} {\bibinfo {author} {\bibfnamefont {R.~J.}\ \bibnamefont
  {Holt}}\ and\ \bibinfo {author} {\bibfnamefont {C.~D.}\ \bibnamefont
  {Roberts}},\ }\href {\doibase 10.1103/RevModPhys.82.2991} {\bibfield
  {journal} {\bibinfo  {journal} {Rev. Mod. Phys.}\ }\textbf {\bibinfo {volume}
  {82}},\ \bibinfo {pages} {2991} (\bibinfo {year} {2010})},\ \Eprint
  {http://arxiv.org/abs/1002.4666} {arXiv:1002.4666 [nucl-th]} \BibitemShut
  {NoStop}%
\bibitem [{\citenamefont {Wang}\ and\ \citenamefont
  {Chen}(2015)}]{Wang:2014lua}%
  \BibitemOpen
  \bibfield  {author} {\bibinfo {author} {\bibfnamefont {R.}~\bibnamefont
  {Wang}}\ and\ \bibinfo {author} {\bibfnamefont {X.}~\bibnamefont {Chen}},\
  }\href {\doibase 10.1103/PhysRevD.91.054026} {\bibfield  {journal} {\bibinfo
  {journal} {Phys. Rev. D}\ }\textbf {\bibinfo {volume} {91}},\ \bibinfo
  {pages} {054026} (\bibinfo {year} {2015})},\ \Eprint
  {http://arxiv.org/abs/1410.3598} {arXiv:1410.3598 [hep-ph]} \BibitemShut
  {NoStop}%
\bibitem [{\citenamefont {Han}\ and\ \citenamefont {Chen}(2017)}]{Han:2016bsw}%
  \BibitemOpen
  \bibfield  {author} {\bibinfo {author} {\bibfnamefont {C.}~\bibnamefont
  {Han}}\ and\ \bibinfo {author} {\bibfnamefont {X.}~\bibnamefont {Chen}},\
  }\href {\doibase 10.1088/1674-1137/41/11/113103} {\bibfield  {journal}
  {\bibinfo  {journal} {Chin. Phys. C}\ }\textbf {\bibinfo {volume} {41}},\
  \bibinfo {pages} {113103} (\bibinfo {year} {2017})},\ \Eprint
  {http://arxiv.org/abs/1612.06631} {arXiv:1612.06631 [hep-ph]} \BibitemShut
  {NoStop}%
\bibitem [{\citenamefont {Han}\ \emph {et~al.}(2020)\citenamefont {Han},
  \citenamefont {Xing}, \citenamefont {Wang}, \citenamefont {Fu}, \citenamefont
  {Wang},\ and\ \citenamefont {Chen}}]{Han:2018wsw}%
  \BibitemOpen
  \bibfield  {author} {\bibinfo {author} {\bibfnamefont {C.}~\bibnamefont
  {Han}}, \bibinfo {author} {\bibfnamefont {H.}~\bibnamefont {Xing}}, \bibinfo
  {author} {\bibfnamefont {X.}~\bibnamefont {Wang}}, \bibinfo {author}
  {\bibfnamefont {Q.}~\bibnamefont {Fu}}, \bibinfo {author} {\bibfnamefont
  {R.}~\bibnamefont {Wang}}, \ and\ \bibinfo {author} {\bibfnamefont
  {X.}~\bibnamefont {Chen}},\ }\href {\doibase 10.1016/j.physletb.2019.135066}
  {\bibfield  {journal} {\bibinfo  {journal} {Phys. Lett. B}\ }\textbf
  {\bibinfo {volume} {800}},\ \bibinfo {pages} {135066} (\bibinfo {year}
  {2020})},\ \Eprint {http://arxiv.org/abs/1809.01549} {arXiv:1809.01549
  [hep-ph]} \BibitemShut {NoStop}%
\bibitem [{\citenamefont {Han}\ \emph {et~al.}(2021)\citenamefont {Han},
  \citenamefont {Xie}, \citenamefont {Wang},\ and\ \citenamefont
  {Chen}}]{Han:2020vjp}%
  \BibitemOpen
  \bibfield  {author} {\bibinfo {author} {\bibfnamefont {C.}~\bibnamefont
  {Han}}, \bibinfo {author} {\bibfnamefont {G.}~\bibnamefont {Xie}}, \bibinfo
  {author} {\bibfnamefont {R.}~\bibnamefont {Wang}}, \ and\ \bibinfo {author}
  {\bibfnamefont {X.}~\bibnamefont {Chen}},\ }\href {\doibase
  10.1140/epjc/s10052-021-09087-8} {\bibfield  {journal} {\bibinfo  {journal}
  {Eur. Phys. J. C}\ }\textbf {\bibinfo {volume} {81}},\ \bibinfo {pages} {302}
  (\bibinfo {year} {2021})},\ \Eprint {http://arxiv.org/abs/2010.14284}
  {arXiv:2010.14284 [hep-ph]} \BibitemShut {NoStop}%
\bibitem [{\citenamefont {Han}\ \emph {et~al.}(2024)\citenamefont {Han},
  \citenamefont {Kou}, \citenamefont {Wang},\ and\ \citenamefont
  {Chen}}]{Han:2024xjg}%
  \BibitemOpen
  \bibfield  {author} {\bibinfo {author} {\bibfnamefont {C.}~\bibnamefont
  {Han}}, \bibinfo {author} {\bibfnamefont {W.}~\bibnamefont {Kou}}, \bibinfo
  {author} {\bibfnamefont {X.}~\bibnamefont {Wang}}, \ and\ \bibinfo {author}
  {\bibfnamefont {X.}~\bibnamefont {Chen}},\ }\href {\doibase
  10.1140/epjc/s10052-024-13397-y} {\bibfield  {journal} {\bibinfo  {journal}
  {Eur. Phys. J. C}\ }\textbf {\bibinfo {volume} {84}},\ \bibinfo {pages}
  {1067} (\bibinfo {year} {2024})},\ \Eprint {http://arxiv.org/abs/2407.05923}
  {arXiv:2407.05923 [hep-ph]} \BibitemShut {NoStop}%
\bibitem [{\citenamefont {Chen}\ \emph {et~al.}(2014)\citenamefont {Chen},
  \citenamefont {Ruan}, \citenamefont {Wang}, \citenamefont {Zhang},\ and\
  \citenamefont {Zhu}}]{Chen:2013nga}%
  \BibitemOpen
  \bibfield  {author} {\bibinfo {author} {\bibfnamefont {X.}~\bibnamefont
  {Chen}}, \bibinfo {author} {\bibfnamefont {J.}~\bibnamefont {Ruan}}, \bibinfo
  {author} {\bibfnamefont {R.}~\bibnamefont {Wang}}, \bibinfo {author}
  {\bibfnamefont {P.}~\bibnamefont {Zhang}}, \ and\ \bibinfo {author}
  {\bibfnamefont {W.}~\bibnamefont {Zhu}},\ }\href {\doibase
  10.1142/S0218301314500578} {\bibfield  {journal} {\bibinfo  {journal} {Int.
  J. Mod. Phys. E}\ }\textbf {\bibinfo {volume} {23}},\ \bibinfo {pages}
  {1450057} (\bibinfo {year} {2014})},\ \Eprint
  {http://arxiv.org/abs/1306.1872} {arXiv:1306.1872 [hep-ph]} \BibitemShut
  {NoStop}%
\bibitem [{\citenamefont {Wang}\ \emph {et~al.}(2017)\citenamefont {Wang},
  \citenamefont {Chen},\ and\ \citenamefont {Fu}}]{Wang:2016mzo}%
  \BibitemOpen
  \bibfield  {author} {\bibinfo {author} {\bibfnamefont {R.}~\bibnamefont
  {Wang}}, \bibinfo {author} {\bibfnamefont {X.}~\bibnamefont {Chen}}, \ and\
  \bibinfo {author} {\bibfnamefont {Q.}~\bibnamefont {Fu}},\ }\href {\doibase
  10.1016/j.nuclphysb.2017.04.008} {\bibfield  {journal} {\bibinfo  {journal}
  {Nucl. Phys. B}\ }\textbf {\bibinfo {volume} {920}},\ \bibinfo {pages} {1}
  (\bibinfo {year} {2017})},\ \Eprint {http://arxiv.org/abs/1611.03670}
  {arXiv:1611.03670 [hep-ph]} \BibitemShut {NoStop}%
\bibitem [{\citenamefont {Wang}\ and\ \citenamefont
  {Chen}(2017)}]{Wang:2016sfq}%
  \BibitemOpen
  \bibfield  {author} {\bibinfo {author} {\bibfnamefont {R.}~\bibnamefont
  {Wang}}\ and\ \bibinfo {author} {\bibfnamefont {X.}~\bibnamefont {Chen}},\
  }\href {\doibase 10.1088/1674-1137/41/5/053103} {\bibfield  {journal}
  {\bibinfo  {journal} {Chin. Phys. C}\ }\textbf {\bibinfo {volume} {41}},\
  \bibinfo {pages} {053103} (\bibinfo {year} {2017})},\ \Eprint
  {http://arxiv.org/abs/1609.01831} {arXiv:1609.01831 [hep-ph]} \BibitemShut
  {NoStop}%
\bibitem [{\citenamefont {Chen}\ \emph {et~al.}(2016)\citenamefont {Chen},
  \citenamefont {Ruan}, \citenamefont {Wang}, \citenamefont {Zhang},\ and\
  \citenamefont {Zhu}}]{Chen:2014nba}%
  \BibitemOpen
  \bibfield  {author} {\bibinfo {author} {\bibfnamefont {X.}~\bibnamefont
  {Chen}}, \bibinfo {author} {\bibfnamefont {J.}~\bibnamefont {Ruan}}, \bibinfo
  {author} {\bibfnamefont {R.}~\bibnamefont {Wang}}, \bibinfo {author}
  {\bibfnamefont {P.}~\bibnamefont {Zhang}}, \ and\ \bibinfo {author}
  {\bibfnamefont {W.}~\bibnamefont {Zhu}},\ }\href {\doibase
  10.1140/epjp/i2016-16006-x} {\bibfield  {journal} {\bibinfo  {journal} {Eur.
  Phys. J. Plus}\ }\textbf {\bibinfo {volume} {131}},\ \bibinfo {pages} {6}
  (\bibinfo {year} {2016})},\ \Eprint {http://arxiv.org/abs/1404.0759}
  {arXiv:1404.0759 [hep-ph]} \BibitemShut {NoStop}%
\bibitem [{\citenamefont {Parisi}\ and\ \citenamefont
  {Petronzio}(1976)}]{Parisi:1976fz}%
  \BibitemOpen
  \bibfield  {author} {\bibinfo {author} {\bibfnamefont {G.}~\bibnamefont
  {Parisi}}\ and\ \bibinfo {author} {\bibfnamefont {R.}~\bibnamefont
  {Petronzio}},\ }\href {\doibase 10.1016/0370-2693(76)90088-5} {\bibfield
  {journal} {\bibinfo  {journal} {Phys. Lett. B}\ }\textbf {\bibinfo {volume}
  {62}},\ \bibinfo {pages} {331} (\bibinfo {year} {1976})}\BibitemShut
  {NoStop}%
\bibitem [{\citenamefont {Vainshtein}\ \emph {et~al.}(1976)\citenamefont
  {Vainshtein}, \citenamefont {Zakharov}, \citenamefont {Novikov},\ and\
  \citenamefont {Shifman}}]{Vainshtein:1976kd}%
  \BibitemOpen
  \bibfield  {author} {\bibinfo {author} {\bibfnamefont {A.~I.}\ \bibnamefont
  {Vainshtein}}, \bibinfo {author} {\bibfnamefont {V.~I.}\ \bibnamefont
  {Zakharov}}, \bibinfo {author} {\bibfnamefont {V.~A.}\ \bibnamefont
  {Novikov}}, \ and\ \bibinfo {author} {\bibfnamefont {M.~A.}\ \bibnamefont
  {Shifman}},\ }\href@noop {} {\bibfield  {journal} {\bibinfo  {journal} {JETP
  Lett.}\ }\textbf {\bibinfo {volume} {24}},\ \bibinfo {pages} {341} (\bibinfo
  {year} {1976})}\BibitemShut {NoStop}%
\bibitem [{\citenamefont {Gluck}\ and\ \citenamefont
  {Reya}(1977)}]{Gluck:1977ah}%
  \BibitemOpen
  \bibfield  {author} {\bibinfo {author} {\bibfnamefont {M.}~\bibnamefont
  {Gluck}}\ and\ \bibinfo {author} {\bibfnamefont {E.}~\bibnamefont {Reya}},\
  }\href {\doibase 10.1016/0550-3213(77)90393-5} {\bibfield  {journal}
  {\bibinfo  {journal} {Nucl. Phys. B}\ }\textbf {\bibinfo {volume} {130}},\
  \bibinfo {pages} {76} (\bibinfo {year} {1977})}\BibitemShut {NoStop}%
\bibitem [{\citenamefont {Gl\"uck}\ \emph {et~al.}(1998)\citenamefont
  {Gl\"uck}, \citenamefont {Reya},\ and\ \citenamefont {Vogt}}]{Gluck:1998xa}%
  \BibitemOpen
  \bibfield  {author} {\bibinfo {author} {\bibfnamefont {M.}~\bibnamefont
  {Gl\"uck}}, \bibinfo {author} {\bibfnamefont {E.}~\bibnamefont {Reya}}, \
  and\ \bibinfo {author} {\bibfnamefont {A.}~\bibnamefont {Vogt}},\ }\href
  {\doibase 10.1007/s100520050289} {\bibfield  {journal} {\bibinfo  {journal}
  {Eur. Phys. J. C}\ }\textbf {\bibinfo {volume} {5}},\ \bibinfo {pages} {461}
  (\bibinfo {year} {1998})},\ \Eprint {http://arxiv.org/abs/hep-ph/9806404}
  {arXiv:hep-ph/9806404} \BibitemShut {NoStop}%
\bibitem [{\citenamefont {Pumplin}\ \emph {et~al.}(2002)\citenamefont
  {Pumplin}, \citenamefont {Stump}, \citenamefont {Huston}, \citenamefont
  {Lai}, \citenamefont {Nadolsky},\ and\ \citenamefont
  {Tung}}]{Pumplin:2002vw}%
  \BibitemOpen
  \bibfield  {author} {\bibinfo {author} {\bibfnamefont {J.}~\bibnamefont
  {Pumplin}}, \bibinfo {author} {\bibfnamefont {D.~R.}\ \bibnamefont {Stump}},
  \bibinfo {author} {\bibfnamefont {J.}~\bibnamefont {Huston}}, \bibinfo
  {author} {\bibfnamefont {H.~L.}\ \bibnamefont {Lai}}, \bibinfo {author}
  {\bibfnamefont {P.~M.}\ \bibnamefont {Nadolsky}}, \ and\ \bibinfo {author}
  {\bibfnamefont {W.~K.}\ \bibnamefont {Tung}},\ }\href {\doibase
  10.1088/1126-6708/2002/07/012} {\bibfield  {journal} {\bibinfo  {journal}
  {JHEP}\ }\textbf {\bibinfo {volume} {07}},\ \bibinfo {pages} {012} (\bibinfo
  {year} {2002})},\ \Eprint {http://arxiv.org/abs/hep-ph/0201195}
  {arXiv:hep-ph/0201195} \BibitemShut {NoStop}%
\bibitem [{\citenamefont {Bloom}\ and\ \citenamefont
  {Gilman}(1971)}]{Bloom:1971ye}%
  \BibitemOpen
  \bibfield  {author} {\bibinfo {author} {\bibfnamefont {E.~D.}\ \bibnamefont
  {Bloom}}\ and\ \bibinfo {author} {\bibfnamefont {F.~J.}\ \bibnamefont
  {Gilman}},\ }\href {\doibase 10.1103/PhysRevD.4.2901} {\bibfield  {journal}
  {\bibinfo  {journal} {Phys. Rev. D}\ }\textbf {\bibinfo {volume} {4}},\
  \bibinfo {pages} {2901} (\bibinfo {year} {1971})}\BibitemShut {NoStop}%
\bibitem [{\citenamefont {Bloom}\ and\ \citenamefont
  {Gilman}(1970)}]{Bloom:1970xb}%
  \BibitemOpen
  \bibfield  {author} {\bibinfo {author} {\bibfnamefont {E.~D.}\ \bibnamefont
  {Bloom}}\ and\ \bibinfo {author} {\bibfnamefont {F.~J.}\ \bibnamefont
  {Gilman}},\ }\href {\doibase 10.1103/PhysRevLett.25.1140} {\bibfield
  {journal} {\bibinfo  {journal} {Phys. Rev. Lett.}\ }\textbf {\bibinfo
  {volume} {25}},\ \bibinfo {pages} {1140} (\bibinfo {year}
  {1970})}\BibitemShut {NoStop}%
\bibitem [{\citenamefont {Wilson}(1974)}]{Wilson:1974sk}%
  \BibitemOpen
  \bibfield  {author} {\bibinfo {author} {\bibfnamefont {K.~G.}\ \bibnamefont
  {Wilson}},\ }\href {\doibase 10.1103/PhysRevD.10.2445} {\bibfield  {journal}
  {\bibinfo  {journal} {Phys. Rev. D}\ }\textbf {\bibinfo {volume} {10}},\
  \bibinfo {pages} {2445} (\bibinfo {year} {1974})}\BibitemShut {NoStop}%
\bibitem [{\citenamefont {Zyla}\ \emph {et~al.}(2020)\citenamefont {Zyla} \emph
  {et~al.}}]{zyla202079}%
  \BibitemOpen
  \bibfield  {author} {\bibinfo {author} {\bibfnamefont {P.}~\bibnamefont
  {Zyla}} \emph {et~al.},\ }\href@noop {} {\bibfield  {journal} {\bibinfo
  {journal} {Prog. Theor. Exp. Phys}\ }\textbf {\bibinfo {volume} {2020}},\
  \bibinfo {pages} {083C01} (\bibinfo {year} {2020})}\BibitemShut {NoStop}%
\bibitem [{\citenamefont {Dokshitzer}(1977)}]{Dokshitzer:1977sg}%
  \BibitemOpen
  \bibfield  {author} {\bibinfo {author} {\bibfnamefont {Y.~L.}\ \bibnamefont
  {Dokshitzer}},\ }\href@noop {} {\bibfield  {journal} {\bibinfo  {journal}
  {Sov. Phys. JETP}\ }\textbf {\bibinfo {volume} {46}},\ \bibinfo {pages} {641}
  (\bibinfo {year} {1977})}\BibitemShut {NoStop}%
\bibitem [{\citenamefont {Gribov}\ and\ \citenamefont
  {Lipatov}(1972)}]{Gribov:1972ri}%
  \BibitemOpen
  \bibfield  {author} {\bibinfo {author} {\bibfnamefont {V.~N.}\ \bibnamefont
  {Gribov}}\ and\ \bibinfo {author} {\bibfnamefont {L.~N.}\ \bibnamefont
  {Lipatov}},\ }\href@noop {} {\bibfield  {journal} {\bibinfo  {journal} {Sov.
  J. Nucl. Phys.}\ }\textbf {\bibinfo {volume} {15}},\ \bibinfo {pages} {438}
  (\bibinfo {year} {1972})}\BibitemShut {NoStop}%
\bibitem [{\citenamefont {Altarelli}\ and\ \citenamefont
  {Parisi}(1977)}]{Altarelli:1977zs}%
  \BibitemOpen
  \bibfield  {author} {\bibinfo {author} {\bibfnamefont {G.}~\bibnamefont
  {Altarelli}}\ and\ \bibinfo {author} {\bibfnamefont {G.}~\bibnamefont
  {Parisi}},\ }\href {\doibase 10.1016/0550-3213(77)90384-4} {\bibfield
  {journal} {\bibinfo  {journal} {Nucl. Phys. B}\ }\textbf {\bibinfo {volume}
  {126}},\ \bibinfo {pages} {298} (\bibinfo {year} {1977})}\BibitemShut
  {NoStop}%
\bibitem [{\citenamefont {Callan}\ and\ \citenamefont
  {Gross}(1969)}]{Callan:1969uq}%
  \BibitemOpen
  \bibfield  {author} {\bibinfo {author} {\bibfnamefont {C.~G.}\ \bibnamefont
  {Callan}, \bibfnamefont {Jr.}}\ and\ \bibinfo {author} {\bibfnamefont
  {D.~J.}\ \bibnamefont {Gross}},\ }\href {\doibase 10.1103/PhysRevLett.22.156}
  {\bibfield  {journal} {\bibinfo  {journal} {Phys. Rev. Lett.}\ }\textbf
  {\bibinfo {volume} {22}},\ \bibinfo {pages} {156} (\bibinfo {year}
  {1969})}\BibitemShut {NoStop}%
\bibitem [{\citenamefont {Farrar}\ and\ \citenamefont
  {Jackson}(1975)}]{Farrar:1975yb}%
  \BibitemOpen
  \bibfield  {author} {\bibinfo {author} {\bibfnamefont {G.~R.}\ \bibnamefont
  {Farrar}}\ and\ \bibinfo {author} {\bibfnamefont {D.~R.}\ \bibnamefont
  {Jackson}},\ }\href {\doibase 10.1103/PhysRevLett.35.1416} {\bibfield
  {journal} {\bibinfo  {journal} {Phys. Rev. Lett.}\ }\textbf {\bibinfo
  {volume} {35}},\ \bibinfo {pages} {1416} (\bibinfo {year}
  {1975})}\BibitemShut {NoStop}%
\bibitem [{\citenamefont {Brodsky}\ \emph {et~al.}(1995)\citenamefont
  {Brodsky}, \citenamefont {Burkardt},\ and\ \citenamefont
  {Schmidt}}]{Brodsky:1994kg}%
  \BibitemOpen
  \bibfield  {author} {\bibinfo {author} {\bibfnamefont {S.~J.}\ \bibnamefont
  {Brodsky}}, \bibinfo {author} {\bibfnamefont {M.}~\bibnamefont {Burkardt}}, \
  and\ \bibinfo {author} {\bibfnamefont {I.}~\bibnamefont {Schmidt}},\ }\href
  {\doibase 10.1016/0550-3213(95)00009-H} {\bibfield  {journal} {\bibinfo
  {journal} {Nucl. Phys. B}\ }\textbf {\bibinfo {volume} {441}},\ \bibinfo
  {pages} {197} (\bibinfo {year} {1995})},\ \Eprint
  {http://arxiv.org/abs/hep-ph/9401328} {arXiv:hep-ph/9401328} \BibitemShut
  {NoStop}%
\bibitem [{\citenamefont {Feynman}(2018)}]{feynman2018photon}%
  \BibitemOpen
  \bibfield  {author} {\bibinfo {author} {\bibfnamefont {R.~P.}\ \bibnamefont
  {Feynman}},\ }\href@noop {} {\emph {\bibinfo {title} {Photon-hadron
  interactions}}}\ (\bibinfo  {publisher} {CRC Press},\ \bibinfo {year}
  {2018})\BibitemShut {NoStop}%
\bibitem [{\citenamefont {Close}(1973)}]{Close:1973xw}%
  \BibitemOpen
  \bibfield  {author} {\bibinfo {author} {\bibfnamefont {F.~E.}\ \bibnamefont
  {Close}},\ }\href {\doibase 10.1016/0370-2693(73)90389-4} {\bibfield
  {journal} {\bibinfo  {journal} {Phys. Lett. B}\ }\textbf {\bibinfo {volume}
  {43}},\ \bibinfo {pages} {422} (\bibinfo {year} {1973})}\BibitemShut
  {NoStop}%
\bibitem [{\citenamefont {Isgur}(1999)}]{Isgur:1998yb}%
  \BibitemOpen
  \bibfield  {author} {\bibinfo {author} {\bibfnamefont {N.}~\bibnamefont
  {Isgur}},\ }\href {\doibase 10.1103/PhysRevD.59.034013} {\bibfield  {journal}
  {\bibinfo  {journal} {Phys. Rev. D}\ }\textbf {\bibinfo {volume} {59}},\
  \bibinfo {pages} {034013} (\bibinfo {year} {1999})},\ \Eprint
  {http://arxiv.org/abs/hep-ph/9809255} {arXiv:hep-ph/9809255} \BibitemShut
  {NoStop}%
\end{thebibliography}%

\end{document}